\documentclass[aip,graphicx]{revtex4-1}

\usepackage{graphicx}
\usepackage{dcolumn}
\usepackage{bm}
\usepackage{gensymb}

\draft 

\begin{document}


\title{Ensemble Flow Reconstruction in the Atmospheric Boundary Layer from Spatially Limited Measurements through Latent Diffusion Models} 



\author{Alex Rybchuk}
 \email{alex.rybchuk@nrel.gov}
\author{Malik Hassanaly}%
\author{Nicholas Hamilton}%
\author{Paula Doubrawa}%
\affiliation{ 
National Renewable Energy Laboratory, Golden, Colorado, USA
}%

\author{Mitchell J. Fulton}
\affiliation{%
Department of Mechanical Engineering, University of Colorado Boulder, Boulder, Colorado, USA
}%

\author{Luis A. Martínez-Tossas}
\affiliation{ 
National Renewable Energy Laboratory, Golden, Colorado, USA
}%


\date{\today}

\begin{abstract}
Due to costs and practical constraints, field campaigns in the atmospheric boundary layer typically only measure a fraction of the atmospheric volume of interest. Machine learning techniques have previously successfully reconstructed unobserved regions of flow in canonical fluid mechanics problems and two-dimensional geophysical flows, but these techniques have not yet been demonstrated in the three-dimensional atmospheric boundary layer. Here, we conduct a numerical analogue of a field campaign with spatially limited measurements using large-eddy simulation. We pose flow reconstruction as an inpainting problem, and reconstruct realistic samples of turbulent, three-dimensional flow with the use of a latent diffusion model. The diffusion model generates physically plausible turbulent structures on larger spatial scales, even when input observations cover less than 1\% of the volume. Through a combination of qualitative visualization and quantitative assessment, we demonstrate that the diffusion model generates meaningfully diverse samples when conditioned on just one observation. These samples successfully serve as initial conditions for a large-eddy simulation code. We find that diffusion models show promise and potential for other applications for other turbulent flow reconstruction problems.


\end{abstract}


\maketitle 

\section{Introduction}
Atmospheric field campaigns aim to characterize the complicated state of the atmosphere by deploying several measurement systems, often supplementing observations with modeling. Field campaigns in the atmospheric boundary layer (ABL, approximately the lowest 1 km), are critical for different research areas, such as wind energy \cite{moriartyAmericanWAKEExperimeNt2020,hamilton2022rotor}, air quality \cite{barad1958project,allwineJointUrban20032006}, and wildland fires \cite{warnekeFireInfluenceRegional2023}. While field campaigns in the ABL typically strive to measure as much of the lower atmosphere as possible, they measure only a small fraction of the atmosphere due to practical constraints and the costs associated with observation systems. Models can be used to fill the unmeasured areas between the spatially limited measurements, so that atmospheric dynamics can be characterized across a wide range of scales. This combination of measurements and models enables what is referred to as flow reconstruction: turning spatially and temporally sparse information into a highly resolved field.

Machine learning has emerged as a powerful technique for turbulent flow reconstruction\cite{buzzicottiDataReconstructionComplex2023,yuDeepLearningFluid2023}, joining the ranks of more traditional approaches such as data assimilation \cite{meldiReducedOrderModel2017,chandramouli4DLargeScale2020,monsEnsemblevariationalAssimilationStatistical2021,wangStateEstimationTurbulent2021,zakiLimitedObservationsState2021} and nudging \cite{dileoniSynchronizationBigData2020,zaunerNudgingbasedDataAssimilation2022}. The flow reconstruction problem can be formulated in a variety of approaches, depending on the available data and choice of machine learning architecture. Perhaps the most common framing is that of a ``super-resolution'' problem (see Fukami et al. (2023) \cite{fukamiSuperresolutionAnalysisMachine2023} for a recent review). In super-resolution, low-resolution data is available, and data is upsampled to higher resolution through the use of an algorithm. Turbulence super-resolution has been carried out through the use of convolutional neural network architectures \cite{fukamiSuperresolutionReconstructionTurbulent2019,liuDeepLearningMethods2020,fukamiMachinelearningbasedSpatiotemporalSuper2021} as well as generative adversarial network (GAN) architectures \cite{xieTempoGANTemporallyCoherent2018,dengSuperresolutionReconstructionTurbulent2019,stengelAdversarialSuperresolutionClimatological2020,yousifHighfidelityReconstructionTurbulent2021,hassanalyAdversarialSamplingUnknown2022,wangDeeplearningbasedSuperresolutionReconstruction2022}. 
In an alternative to super-resolution, turbulent flow reconstruction has been posed as an ``inpainting'' problem \cite{buzzicottiReconstructionTurbulentData2021}. In this scenario,  part of an image is masked, and an algorithm plausibly reconstructs the missing data. 
Others have specifically examined the problem of reconstruction given sparse measurements, often through the use of convolutional neural networks \cite{fukamiGlobalFieldReconstruction2021,gundersenSemiconditionalVariationalAutoencoder2021}. Finally, instead of using machine learning architectures rooted in the field of computer vision, others have drawn inspiration from the intersection of partial differential equations and machine learning. Flow reconstruction approaches in these categories employ architectures such as ``physics-informed neural networks'' (PINNs)\cite{caiFlowEspressoCup2021,clarkdileoniReconstructingTurbulentVelocity2023,duStateEstimationMinimal2022} and Deep Operator Networks\cite{clarkdileoniNeuralOperatorPrediction2023}. All in all, the aforementioned flow reconstruction techniques have been applied to a variety of turbulent flow environments, ranging from two-dimensional\cite{dengSuperresolutionReconstructionTurbulent2019} and three-dimensional\cite{yousifDeeplearningApproachReconstructing2023} canonical fluid mechanics problems (e.g., flow behind a cylinder) to two-dimensional geophysical problems\cite{fukamiGlobalFieldReconstruction2021} (e.g., the state of the sea surface).



While flow reconstruction based on machine learning has been shown to be powerful, these techniques have not yet been applied to real-world, three-dimensional geophysical flows, such as an ABL. Real-world, ABL flow reconstruction comes with important challenges that must be accounted for. For example, many observation systems (e.g., Doppler lidar) measure line-of-sight velocity, which often serves as an accurate proxy for just one velocity component (e.g., streamwise velocity $u$) instead of measuring all three velocity components ($u, v, w$). So far, only PINNs have demonstrated the ability to reconstruct all velocity components from measurements of just a scalar field\cite{caiFlowEspressoCup2021}. While PINNs show great potential, they often struggle for multi-scale problems\cite{liPhysicsInformedNeuralOperator2023} such as high $Re$ three-dimensional turbulence, and their success has not yet been demonstrated in an ABL. Another issue in reconstructing real-world flows is the sparseness of ABL measurements, as mentioned earlier. The atmosphere is highly chaotic and flow reconstruction from sparse measurements is typically an ill-posed problem, as many non-unique states could correspond to the same observation. As such, geoscientists often characterize its state probabilistically (e.g., in ensemble-based weather forecasting\cite{wilksStatisticalMethodsAtmospheric2019}). However, most flow reconstruction studies have been practically deterministic (e.g., using GANs, which suffer from the ``mode collapse'' problem \cite{salimansImprovedTechniquesTraining2016}). Only two studies thus far, \citet[][]{gundersenSemiconditionalVariationalAutoencoder2021} using variational autoencoders and \citet[][]{hassanalyAdversarialSamplingUnknown2022} using modified GANs, have reconstructed turbulent flow in a probabilistic manner, though it is not clear if their two-dimensional architectures can computationally scale to a three-dimensional, highly turbulent ABL.

Recently, a new neural architecture known as a diffusion model (DM) \cite{sohl-dicksteinDeepUnsupervisedLearning2015,songGenerativeModelingEstimating2020} has achieved state-of-the-art status for generating high-resolution imagery \cite{dhariwalDiffusionModelsBeat2021}, showing promise for turbulent flow reconstruction in the ABL. A specific category of DM, known as a latent diffusion model (LDM) \cite{rombachHighResolutionImageSynthesis2022} is computationally efficient, and LDMs have been used to generate high-resolution two-dimensional imagery \cite{rombachHighResolutionImageSynthesis2022} and three-dimensional medical imagery \cite{pinayaBrainImagingGeneration2022}. DMs are inherently stochastic, and they have been shown to generate diverse, photorealistic imagery when a random seed is changed. Finally, while DMs have shown strong performance for the super-resolution problem \cite{rombachHighResolutionImageSynthesis2022}, they have also excelled at inpainting \cite{lugmayrRePaintInpaintingUsing2022,sahariaPaletteImagetoImageDiffusion2022}. We posit that the inpainting perspective is a natural approach to pose the problem of flow reconstruction from spatially limited measurements.

In this paper, we ask: given measurements in one small region of the atmosphere, can we estimate the instantaneous, unmeasured state of the ABL nearby through inpainting with an LDM? Here, we study this problem in the context of a synthetic field campaign that is conducted through large-eddy simulation (LES). In this work, our flow reconstruction strategy is to recreate instantaneous, volumetric flow fields. This strategy shares similarities with data assimilation approaches that are commonly applied for mesoscale and synoptic scale atmospheric reconstruction, namely 4DVAR\cite{courtierStrategyOperationalImplementation1994} and the ensemble Kalman filter \cite{evensenSequentialDataAssimilation1994}, which produce plausible initial conditions for a dynamical solver given a time-history of observations. Our investigation here addresses three key ABL reconstruction challenges in a synthetic environment: (1) probabilistic reconstruction, (2) dealing with spatially limited measurements, and (3) translating from scalar measurements to all three velocity components. For the time being, we omit two additional challenges of ABL reconstruction, namely dealing with (1) noisy measurements and (2) a time-history of measurements. An optimal ABL reconstruction strategy would likely be able to account for all these challenges.

We find that LDMs can successfully reconstruct ABL flow given limited measurements, and we believe that LDMs can be applied for other turbulent flow reconstruction problems as well. In this paper, we demonstrate the following contributions:
\begin{itemize}
    \item Through the use of LDMs, we can generate diverse three-dimensional turbulent flow fields. We characterize the quality of LDM samples through the use of qualitative visualization and quantitative assessment. 
    \item We show that LDMs can reconstruct all three velocity components ($u$, $v$, $w$), even when given observations of almost exclusively $u$.
    \item While LDM studies in the field of computer vision have masked upwards of 75\% of an image, we show that LDM reconstructions can be conditioned, even when minimal observations ($<$1\% of the volume) are provided.
    \item We demonstrate that a LDM sample can successfully be used as an initial condition for an LES. To our knowledge, this is the first time that a machine learning sample has demonstrated this kind of compatibility with an LES code.
\end{itemize}

The rest of this manuscript is structured as follows. In Section \ref{sec:data}, we describe our synthetic field campaign, the configuration of the flow reconstruction problem, and the LES dataset that is used to train and test the LDM. In Section \ref{sec:methods}, we provide details of the LDM architecture. In Section \ref{sec:results}, we study the performance of the LDMs. Finally, in Section \ref{sec:conclusion}, we conclude the paper and discuss potential future lines of inquiry.

\section{Data}
\label{sec:data}
In this work, we explore flow reconstruction in the context of the real-world Rotor Aerodynamics Aeroelastics and Wakes (RAAW) field campaign \citep{hamilton2022rotor}, which seeks to thoroughly characterize the behavior of a single, utility-scale wind turbine with respect to the incoming atmospheric inflow, particularly on shorter timescales (1 second to 10 minutes). Here, we conduct a synthetic version of the field campaign, which is sometimes referred to as an ``observing system simulation experiment.'' 

We simulate a thermally neutral atmospheric boundary layer using the LES code AMR-Wind \citep{sprague2021exawind}. AMR-Wind solves the incompressible Navier Stokes equations using a finite-volume approach with second-order accuracy in space and time. Details of the spatial and temporal discretization can be found in Almgren et al. (1998) \citep{almgrenConservativeAdaptiveProjection1998}. The simulation is run with a Smagorinsky subgrid-scale model \cite{smagorinskyGENERALCIRCULATIONEXPERIMENTS1963}. The domain is forced by $U =$ 10 m s$^{-1}$ geostrophic winds in the $x$-direction, as well as Coriolis forcing at a latitude of 90$\degree$. The LES domain is sized $(x, y, z) = (1920, 1920, 960)$ m with $(nx, ny, nz) = (128, 128, 64)$ grid points, leading to 15-m resolution in all three directions. The simulation uses a fixed 0.5-s timestep. The LES is initially run for 6 hours to allow turbulence to spin up and become stationary. Afterward, we generate data for the training set by running the LES for 3.5 days of simulated time and saving the $(u, v, w)$ fields everywhere in the domain every minute, totaling 5,040 samples. After the training set is generated, we generate a test dataset by simulating another 3.5 days after the end of the training dataset period. By using test data from after the training period, we ensure there is no cross-contamination between datasets while also achieving good statistical agreement between the two datasets in our heights of interest, rotor span heights of 56.5--183.5 m (Section \ref{sec:results}).

\begin{figure}
\centering
\includegraphics[width=6.0in]{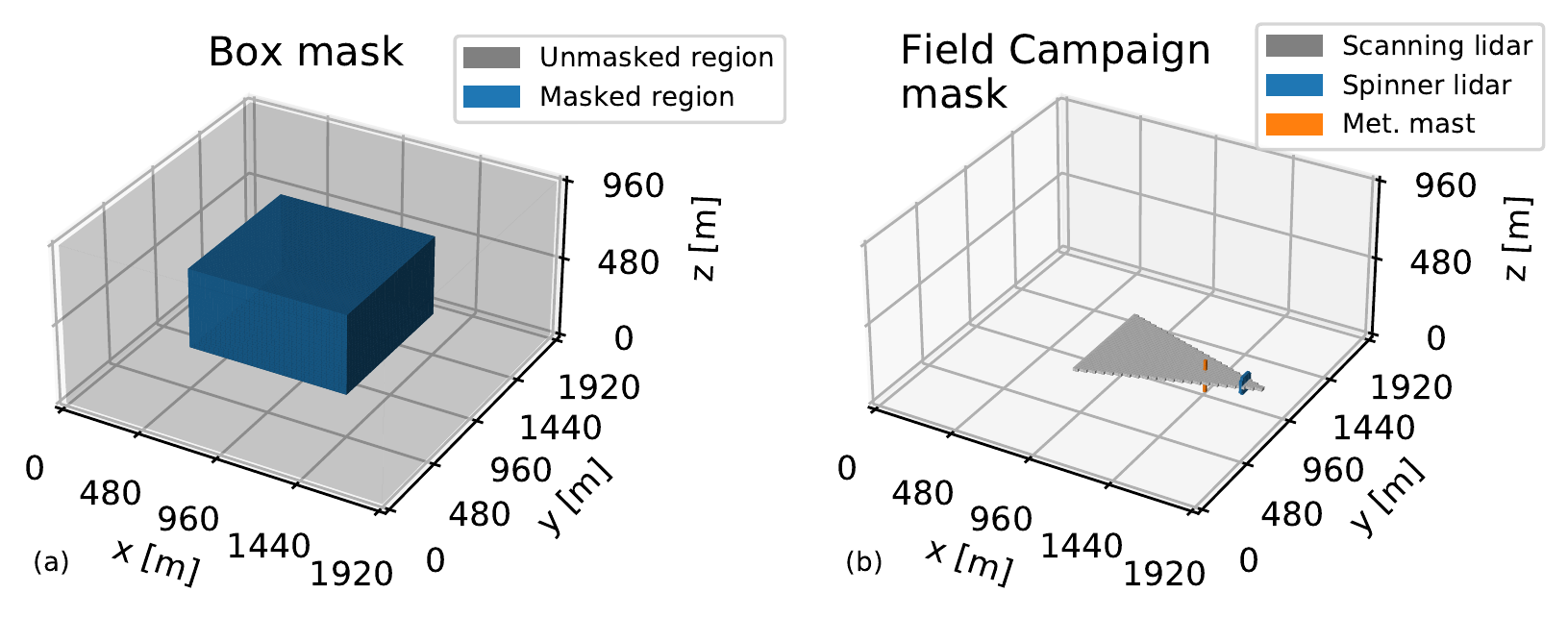}
\caption{(a) The masked region and observed region associated with the box mask. (b) The observations associated with the Field Campaign (FC) mask, with each instrument highlighted. The regions without observations are masked.}
\label{fig:masks}
\end{figure}

Synthetic measurements are generated by masking regions of the three-dimensional LES output, and in the context of this paper, we refer to unmasked regions as ``observations.'' We test the LDM on two sets of masks (Fig. \ref{fig:masks}). The ``box mask'' is commonly used in two-dimensional image inpainting problems \citep{sahariaPaletteImagetoImageDiffusion2022,lugmayrRePaintInpaintingUsing2022}. Here, we mask a cube at the center of the domain spanning $(64,64,32)$ grid points, in order to qualitatively demonstrate flow reconstruction in a scenario with many observations. For the box mask, we use observations of all three components velocity. The ``field campaign (FC) mask'' hides data at locations specific to the RAAW field campaign instrumentation layout, observing data at 1,525 of 1,048,576 LES grid cells. This mask could be modified for any other layout of field measurements.  All instruments are aligned to point into the incoming wind, as they would be in the field. The turbine can be thought of as sitting at $(x,y) = (1800, 960)$ m. The horizontal scanning lidar reaches 1,000 m upwind of the turbine, covering an azimuthal range of 18$\degree$ at a height of 120 m. The vertical spinner lidar scan covers the rotor area and sits 120 m upwind of the turbine. The meteorological mast sits 360 m upwind of the turbine and measures in a vertical column between 0 m and 180 m. 

In this study, we take steps towards developing an algorithm that could be applied to real-world measurements, but we still make a number of simplifying assumptions that will be relaxed in future work.
\begin{itemize}
    \item We assume that measurements are noise-free. 
    \item We assume that both lidars directly measure the $u$-component of velocity and only this component of velocity. In practice, lidars measure line-of-sight velocity, which is then projected to other directions to obtain specific components of the velocity vector. As our lidars measure directly upstream into the oncoming flow, we believe this assumption is warranted. The meteorological mast measures all three components of velocity.
    \item We assume that lidars instantaneously scan their 2D $x-y$ plane of interest. In practice, both lidars scan across their 2D planes in approximately 2–5 seconds.
\end{itemize}

\section{Methods}
\label{sec:methods}

\subsection{Background}

\begin{figure*}[htbp]
\centering
\includegraphics[width=6.0in]{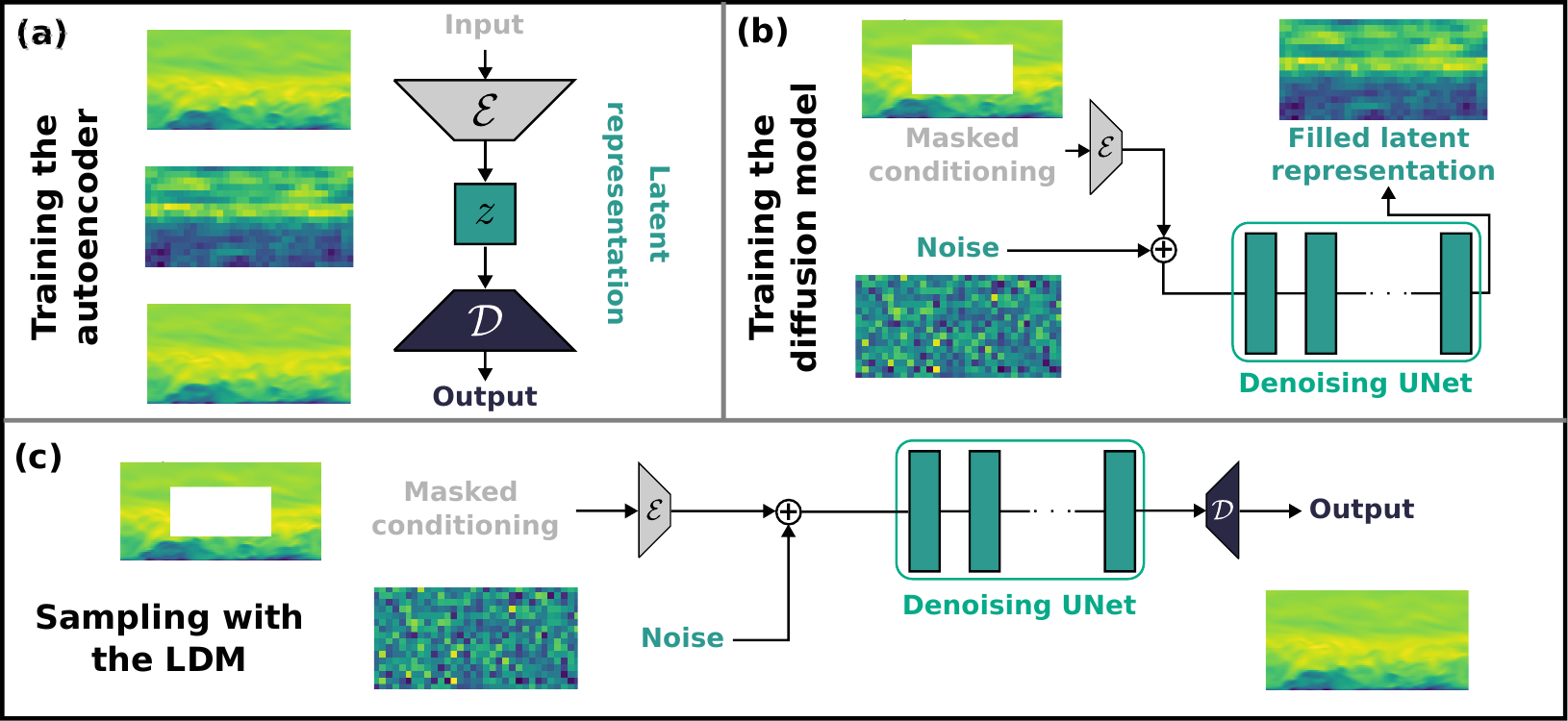}
\caption{A schematic depicting the major components of an LDM and their function. (a) First, the encoder and decoder of an autoencoder are trained. (b) Next, a diffusion model is trained with the help of the encoder from the autoencoder. The conditioning is optional and is not included for unconditional networks. (c) The trained encoder, diffusion model, and decoder are combined into an LDM and used to generate samples, possibly given optional conditioning. Additional details can be found in Fig. \ref{fig:detailed_arch} and Table \ref{table:detailed_arch}.}
\label{fig:ldm_schematic}
\end{figure*}

In order to generate synthetic atmospheric states, we employ an LDM \cite[][Figs. \ref{fig:ldm_schematic}, \ref{fig:detailed_arch}, Table \ref{table:detailed_arch}]{rombachHighResolutionImageSynthesis2022}, an architecture based on DMs. In recent years, DMs have emerged as a new category of deep generative model. They were originally developed through the perspective of nonequilibrium statistical physics \cite{sohl-dicksteinDeepUnsupervisedLearning2015}, but they can be derived through the lens of score-based modeling \cite{songGenerativeModelingEstimating2020} or Markovian hierarchical variational autoencoders \cite{luoUnderstandingDiffusionModels2022}. From one perspective \cite{songScoreBasedGenerativeModeling2021}, DMs are trained with the assistance of a prescribed degradation process, in which Gaussian noise is repeatedly added to a sample $\bm{x}$ over an interval $t \in [0,T]$ until the final sample is indistinguishable from pure Gaussian noise. This process can be thought of as a forward stochastic differential equation

\begin{equation}
    \text{d}\bm{x} = \bm{f}(\bm{x},t) \text{d}t + g(t) \text{d}\bm{w}
\end{equation}

\noindent where $\bm{f}$ is the drift coefficient, $g$ is the diffusion coefficient, and $\bm{w}$ is the standard Wiener process. This process can be undone such that the final Gaussian noise state can be reverted to the initial sample using the reverse stochastic differential equation

\begin{equation}
    \text{d}\bm{x} = [\bm{f}(\bm{x},t) \text{d}t - g(t)^2 \nabla_{\bm{x}} \log p_t(\bm{x})]\text{d}t + g(t) \text{d}\bm{\bar{w}}
\end{equation}

\noindent where $\bm{\bar{w}}$ is the standard Wiener process for the reverse equation and $\nabla \log p_t(\bm{x})$ is the score function, which is challenging to estimate. DMs learn the score function, after which they can be used to draw samples from the probability distribution function $p_0(x)$ (which we will simply refer to as $p(x)$ from here on) that characterizes the training dataset by initially starting with Gaussian noise. Given supplemental information, such as observations or class labels, DMs can draw samples from a conditional probability density function $p(x|y)$ through a number of approaches---for example, training on paired data to directly learn to sample from $p(x|y)$ as is done by \citet{songGenerativeModelingEstimating2020}, or by first learning to sample from the unconditional $p(x)$ and then learning an additional conditioning mechanism \cite[e.g.,][]{kawarDenoisingDiffusionRestoration2022}. We employ the first approach in this study.

While traditional DMs have been shown to generate high-quality images, they can be prohibitively expensive to train for large samples, and three-dimensional samples are often large. LDMs address this issue by compressing raw, pixel-space samples into a latent space with the use of an autoencoder (AE) (Fig. \ref{fig:ldm_schematic}a). The AE is traditionally trained with a mixture of $L_1$ or $L_2$ losses, Kullback–Leibler divergence losses as in a variational AE \cite{kingmaAutoEncodingVariationalBayes2013}, perceptual losses based on the VGG network \cite{zhangUnreasonableEffectivenessDeep2018}, and patch-based adversarial losses \cite{isolaImagetoImageTranslationConditional2018}. After the AE is trained, a diffusion model is trained to generate samples in latent space (Fig. \ref{fig:ldm_schematic}b). After the components of the autoencoder and diffusion model are trained, they can be combined to generate samples in pixel-space (Fig. \ref{fig:ldm_schematic}c).

\subsection{Modifications to the original LDM}
\label{sec:ldm_mods}
We modify the original LDM architecture so that it can be applied to generate synthetic LES data. The exact implementation of the architecture can be found at 

\noindent \href{https://github.com/rybchuk/latent-diffusion-3d-atmospheric-boundary-layer}{https://github.com/rybchuk/latent-diffusion-3d-atmospheric-boundary-layer}, and we provide a summary of the modifications to the LDM code here. 
\begin{itemize}
    \item The original LDM was developed for 2D images, so we modify the architecture to work on 3D data through the use of operations like 3D convolutions and 3D normalization.
    \item While the original LDM uses several attention blocks \cite{brockLargeScaleGAN2019} throughout the AE, we omit their use there due to the computational demands of a 3D attention block.
    \item We replace group normalization \cite{wuGroupNormalization2018} with instance normalization \cite{ulyanovInstanceNormalizationMissing2017} everywhere except within attention blocks, as we found the latter performed better.
    \item  In the process of selecting the finalized architecture for this paper, we experimented with different weights for each of the loss components in the AE. We also added the option for a physics-based loss that assesses mass conservation by calculating the divergence of the velocity field. In the end, we found the best performance by using an $L_1$ term, Kullback–Leibler divergence term, and whole-sample adversarial term in the loss function. We omit the VGG-based adversarial loss as well as the mass conservation loss.
    \item In the original LDM architecture \cite{rombachHighResolutionImageSynthesis2022}, image inpainting is accomplished by first drawing a sample from $p(x|y)$ using unmasked regions as conditioning information. In this scenario, there is no guarantee that the generated sample exactly agrees with the conditioning information $y$, so inpainting is achieved by overlaying $y$ onto the sample in a postprocessing step. In effect, this treats conditioning information as a hard constraint, meaning the observation is exactly matched in the reconstruction, which makes sense when inpainting images. However, we found this hard constraint would lead to artifacts at the boundaries of FC mask observations, likely because these observations are only one pixel wide in certain dimensions. As such, we do not postprocess LDM output to exactly match the observations, thereby treating observations as soft constraints, meaning observations are not exactly matched in reconstructions. In other common atmospheric reconstruction techniques, namely data assimilation \cite{carrassiDataAssimilationGeosciences2018}, observations are also treated as soft constraints. Though, we note that in data assimilation, the relative influence of observations on reconstructions can be controlled by prescribing a certain measurement noise magnitude, whereas in our LDM technique here, we cannot modulate the influence of observations.
\end{itemize}

\subsection{Network configurations}
\begin{figure*}[htbp]
\centering
\includegraphics{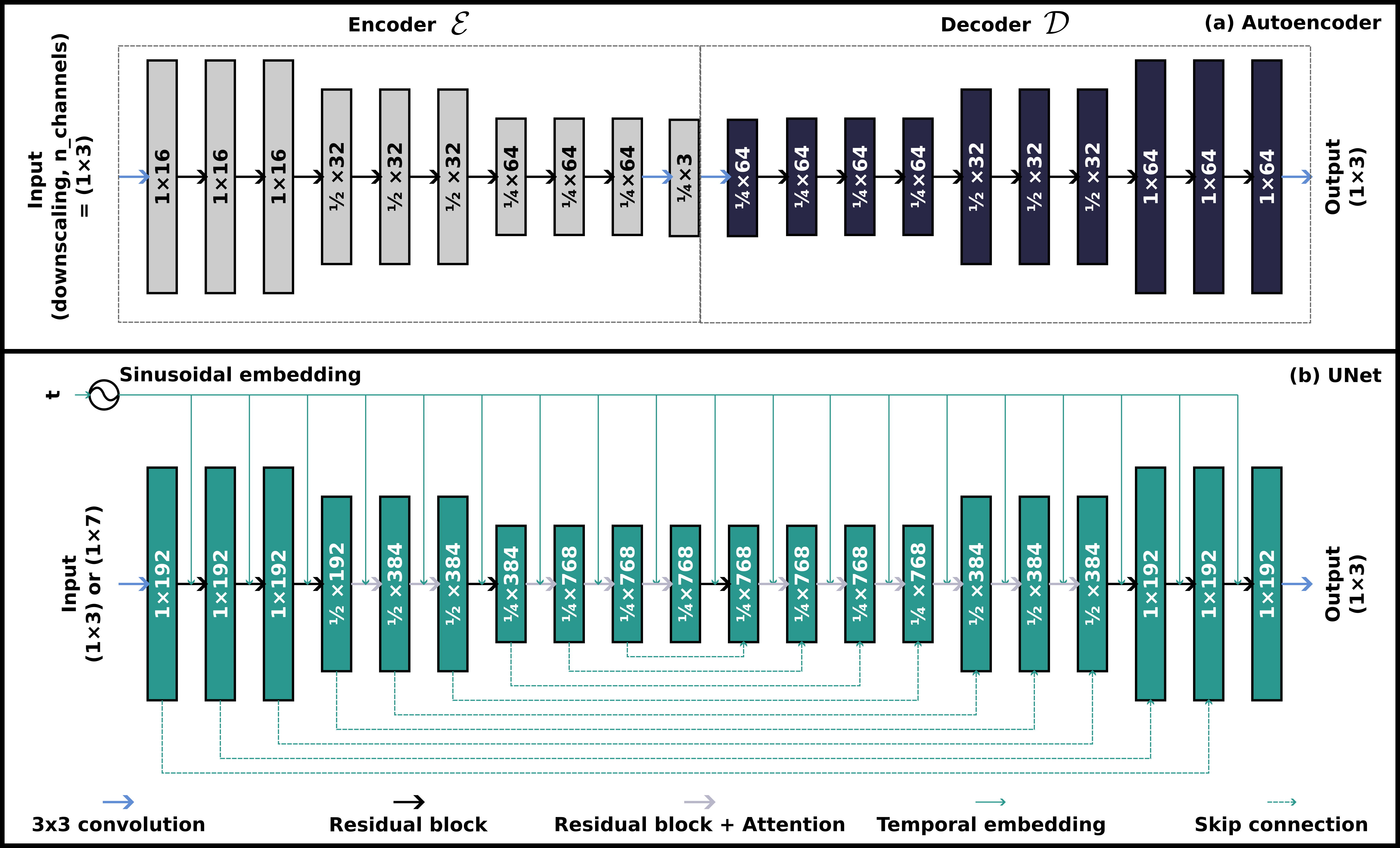}
\caption{A schematic showing the internal components of the LDM in greater detail, including (a) in the autoencoder and (b) in the UNet of the diffusion model.}
\label{fig:detailed_arch}
\end{figure*}

\begin{table}[]
\caption{\label{} Additional parameters used during training of the autoencoder and the diffusion model.}
\begin{tabular}{ll}
\hline
Parameter                       & Value  \\ \hline
Batch size, AE         & 2 \\
Batch size, DM     & 2 \\
Base learning rate, AE & 4.5e-6 \\
Base learning rate, DM  & 5e-5   \\
Noise schedule                  & Linear \\
$N$ noise steps                   & 1000   \\
Start variance                  & 0.0015 \\
End variance                    & 0.0155 \\
$N$ attention heads               & 8      \\ \hline
\end{tabular}
\label{table:detailed_arch}
\end{table}

The LDM requires an AE network for compression to the latent space and decompression (Fig. \ref{fig:detailed_arch}a). Here, the AE has three input, latent space, and output channels that correspond to the velocity variables $(u,v,w)$. Future work could expand this to learn other quantities like temperature, though we prioritize the three velocity components in this study as these have the greatest relevance to wind turbine dynamics. The encoder and decoder of the AE have three internal levels, each with three residual blocks \cite{he2016deep}, compressing data with spatial dimensions of $(128, 128, 64)$ to $(32, 32, 16)$. The first internal level uses 16 channels, followed by 32 channels, and then 64 channels. We use a batch size of 2 due to the high memory use of 3D convolutions. This network is much smaller than the original LDM used in \citet[][https://github.com/CompVis/latent-diffusion]{rombachHighResolutionImageSynthesis2022}, which, for example, had six internal layers with 128 base channels. Our network is smaller due to computational constraints (we train on two 16-GB VRAM V100s) as well as our large three-dimensional samples. Following common practice, we rescale the input data so that the values in each channel fall in the range of -1 to 1.

While our primary goal is to generate candidate velocity fields given FC-style observations, we train both a conditional DM and an unconditional DM to provide more context on network behavior. Both DMs uses the same AE for conversion between pixel space and latent space. The DMs use a UNet architecture \citep{ronnebergerUNetConvolutionalNetworks2015} with three internal layers, respectively using $(192, 384, 768)$ channels, and two residual blocks per layer. Attention blocks are used in the 384 channel layers, 768 channel layers, and the center of the UNet. The DM noise schedule uses 1,000 diffusion steps and is linear. The only difference between the unconditional and conditional DM is the input layer---the unconditional network uses three channels, and the conditional network has an additional four channels, three of which correspond to the compressed observation in latent space and the last of which corresponds to a compressed binary mask corresponding to observed pixels. The conditional DM is exposed to several different masks during training, including box-style and the FC-style observations, in order to increase robustness and potentially accuracy \citep{sahariaPaletteImagetoImageDiffusion2022}. The same conditional DM is then used to produce both box-style and FC-style samples.

\section{Results}
Below, we assess the ability of the LDM to generate plausible atmospheric states. We begin by visualizing unconditional and conditional LDM samples and qualitatively assessing them, as is common in both computer vision manuscripts \citep{songScoreBasedGenerativeModeling2021,rombachHighResolutionImageSynthesis2022} as well as turbulent flow manuscripts \citep{fukamiSuperresolutionReconstructionTurbulent2019,fukamiMachinelearningbasedSpatiotemporalSuper2021,leinonenLatentDiffusionModels2023}. Next, we quantitatively assess the LDM samples by calculating mean profiles, statistics, spectra, and divergence. Finally, we quantitatively assess the ability of the LDM to generate diverse samples from a single observation. 

\label{sec:results}
\subsection{Qualitative assessment of LDM samples}
\label{sec:qual_analysis}
\begin{figure*}[htbp]
\centering
\includegraphics[width=6.3in]{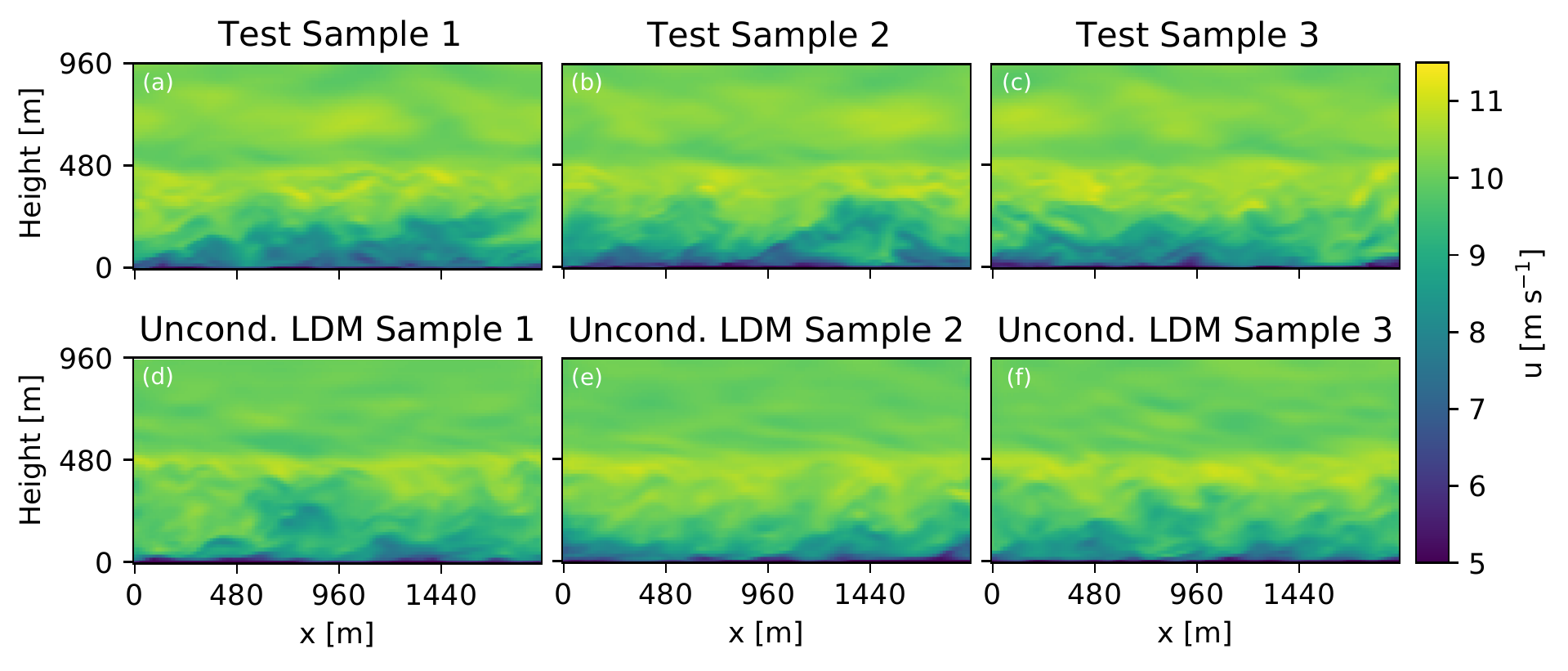}
\caption{(a)--(c) Streamwise cross sections of $u$ at $y$ = $960$ m from ground truth, test data. (d)--(f) Same as above, except for samples from an unconditional LDM.}
\label{fig:qual_uncond_planes}
\end{figure*}


First, in order to broadly demonstrate that LDMs can generate samples that look like an ABL, we compare cross sections of $u$ from the unconditional LDM to those from the ground-truth ABL training data (Fig. \ref{fig:qual_uncond_planes}). Additional qualitative visualizations of $v$ and $w$ as well as three-dimensional isocontours are available in Appendix A. Similar to the training data, the LDM samples have winds that are generally slow near the ground and stronger aloft until they hit the capping inversion near 480 m, above which the winds are less turbulent. The LDM turbulent structures look like turbulent structures in the ground truth data. The horizontal boundaries of the samples are periodic, just as they are in the testing data. Crucially, however, closer inspection reveals that the LDM samples appear slightly smoother than the ground truth data. This discrepancy on smaller scales is further characterized in Section \ref{sec:les_analysis}.

\begin{figure*}[htbp]
\centering
\includegraphics[width=6.3in]{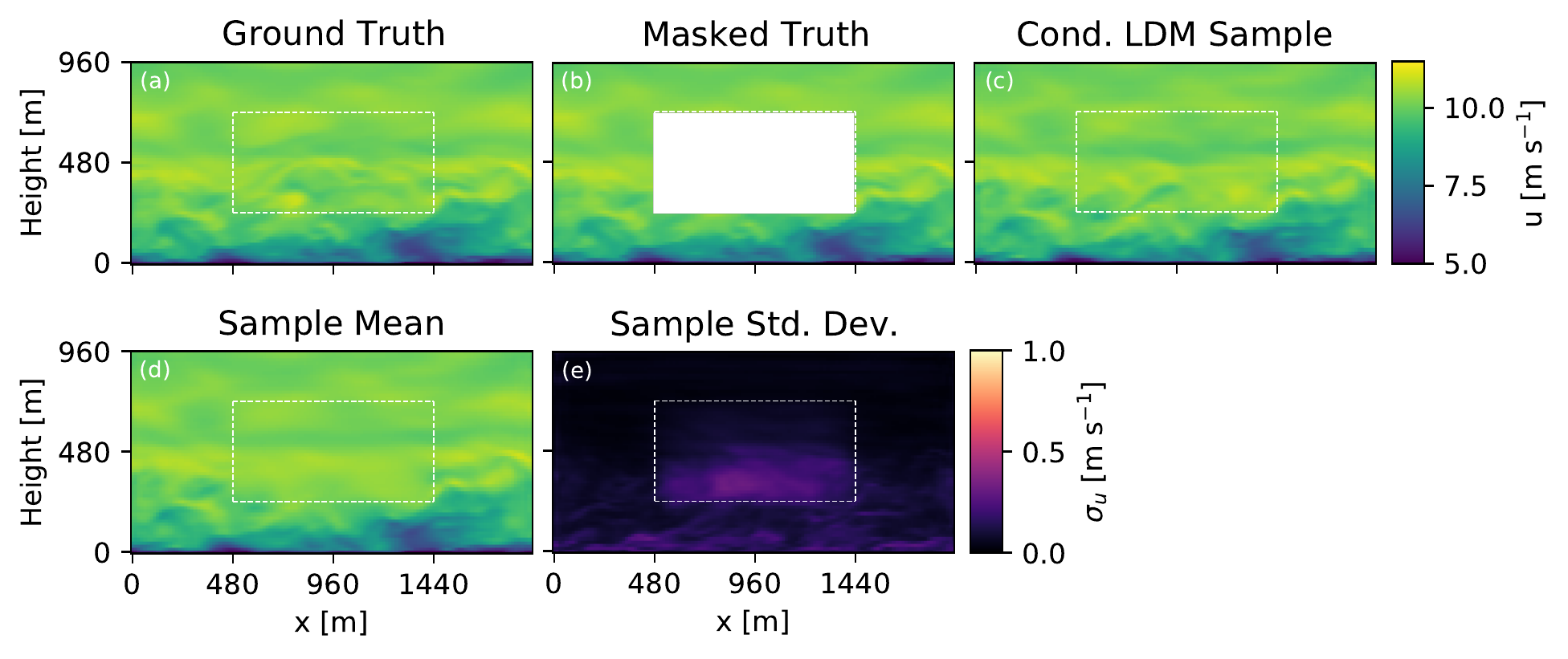}
\caption{Streamwise cross sections of $u$ at $y$ = $960$ m for (a) a ground truth sample, (b) the box observation created from the ground truth sample, and (c) one sample from the conditional LDM, given the observation. (d) The mean from $n=100$ conditional LDM samples. (e) The standard deviation from $n=100$ conditional LDM samples. All subpanels display a white dashed line as a reference to the mask.}
\label{fig:qual_cond_box_planes}
\end{figure*}

We next show that conditional LDMs can inpaint well when provided abundant data, as is the scenario with the box max (Fig. \ref{fig:qual_cond_box_planes}). We visualize a $u$ cross section of a ground truth sample, the associated box mask observation, one conditionally generated sample, and the mean and standard deviation from $n=100$ conditionally generated samples. The conditional samples look similar to the unconditional samples, except the perimeter of the conditional sample and the averaged field is relatively unchanged. The transition between the unmasked and the masked region is smooth and does not display any artifacts. This smooth transition is in part achieved by treating observations as a soft constraint instead of a traditional inpainting hard constraint, in which the exact values of the observations would be superimposed over the unmasked region. The statistical behavior of the conditional predictions qualitatively matches expected behavior. The mean prediction clearly displays the capping inversion, above which the mean flow structures inside the masked region agree well with the structures in the unmasked region. Beneath the capping inversion, the mean prediction is smoothed out when compared to individual turbulent realizations. Inside the unmasked region, the standard deviation of LDM predictions is largest at the bottom of the mask and smaller aloft. This aligns with physical behavior and is expected, as mean shear gradients at the surface encourage mechanical generation of turbulent kinetic energy and therefore wind speed variance. The LDM also displays small, nonzero standard deviation in the unmasked regions, which would not be the case if we treated observations as a hard constraint. 

\begin{figure*}[htbp]
\centering
\includegraphics[width=6.3in]{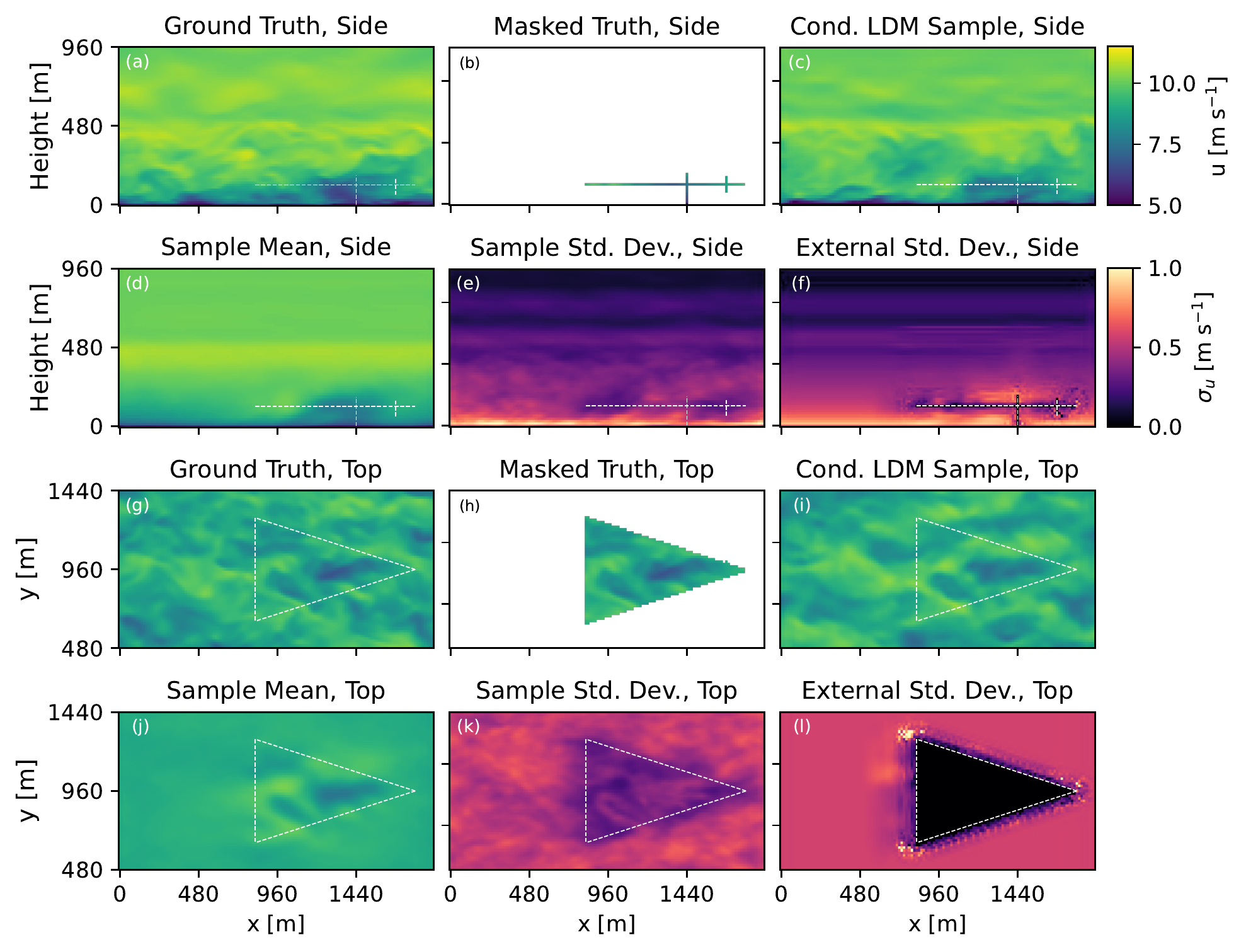}
\caption{(a)--(e) Same as Fig. \ref{fig:qual_cond_box_planes}, except for the FC mask. (f) An external estimate of the conditional standard deviation for the observation in panel (b), which is discussed in greater detail in Section \ref{sec:ens_assessment}. (g)--(l) Corresponding top-down views at the height of the scanning lidar for the data in panels (a)--(f).}
\label{fig:qual_cond_raaw_planes}
\end{figure*}

Finally, we find that conditional LDMs can generate samples that look realistic even when provided minimal observations, as is the case for the FC mask (Fig. \ref{fig:qual_cond_raaw_planes}). We include additional visualizations, including visualizations of $v$ and $w$ which are only conditioned on meteorological mast measurements, in Appendix B. For the particular observation shown in Fig. \ref{fig:qual_cond_raaw_planes}, there is a region of slow $u$-wind in the vicinity of the meteorological mast. Correspondingly, each of the samples also has a patch of slow wind in the same area. This slow flow structure is especially clear in the mean prediction (Fig. \ref{fig:qual_cond_raaw_planes}d, j), and qualitatively, the mean prediction aligns well with the ground truth in the vicinity of the observations. This behavior illustrates that the LDM includes the conditional information of the measurements when generating samples. Away from the observations, the mean prediction becomes much smoother, showing that the conditioning only has local effects, as would be expected in a turbulent flow. The standard deviation of the predictions for the FC mask (Fig. \ref{fig:qual_cond_raaw_planes}e, k) also show a smaller standard deviation in the vicinity of the observations, especially near the horizontal lidar measurement. Thus, the conditioning is reducing the sample-to-sample spread near the measurements, as desired. The FC samples exhibit much higher variance than the box mask samples (Fig. \ref{fig:qual_cond_box_planes}e), as would be expected because the FC samples are conditioned on a smaller region. As we discuss in Section \ref{sec:ens_assessment} in greater detail, the sample standard deviation qualitatively matches up with a different estimate of the standard deviation that comes from another tool (Fig. \ref{fig:qual_cond_box_planes}f).

After assessing the visual quality of samples, we further demonstrate the quality of LDM samples by showing that they can successfully act initial conditions for our LES code. We generate a FC mask sample using the observation in Fig. \ref{fig:qual_cond_raaw_planes} and then use it as the initial condition in a 1-hour LES simulation. We create a video that visualizes a time history of a streamwise cross section and a top-down view of the simulation at https://github.com/rybchuk/latent-diffusion-3d-atmospheric-boundary-layer. The simulation successfully runs and does not display any obvious numerical artifacts such as discontinuities in the flow field around the masked region or ringing in the form of waves or instabilities.


\subsection{Quantitative assessment of LDM samples}
\label{sec:les_analysis}
We next quantitatively assess the performance of the LDMs against physical aspects of the LES: vertical profiles of velocity components and kinematic fluxes, probability density functions of velocity components, turbulence spectra, and mass conservation. For completeness, we compare LDM sample statistics to those from both the training dataset and testing dataset. The training and testing datasets have similar but distinct statistics, as they come from distinct simulation periods. However, as we demonstrate below, their statistics match within our region of interest (rotor disk heights), thereby improving the probability of success for our flow reconstruction problem. To facilitate these quantitative comparisons, we examine characteristics between heights of 0 m to 480 m, namely below the capping inversion.

\begin{figure*}[htbp]
\centering
\includegraphics[width=6.3in]{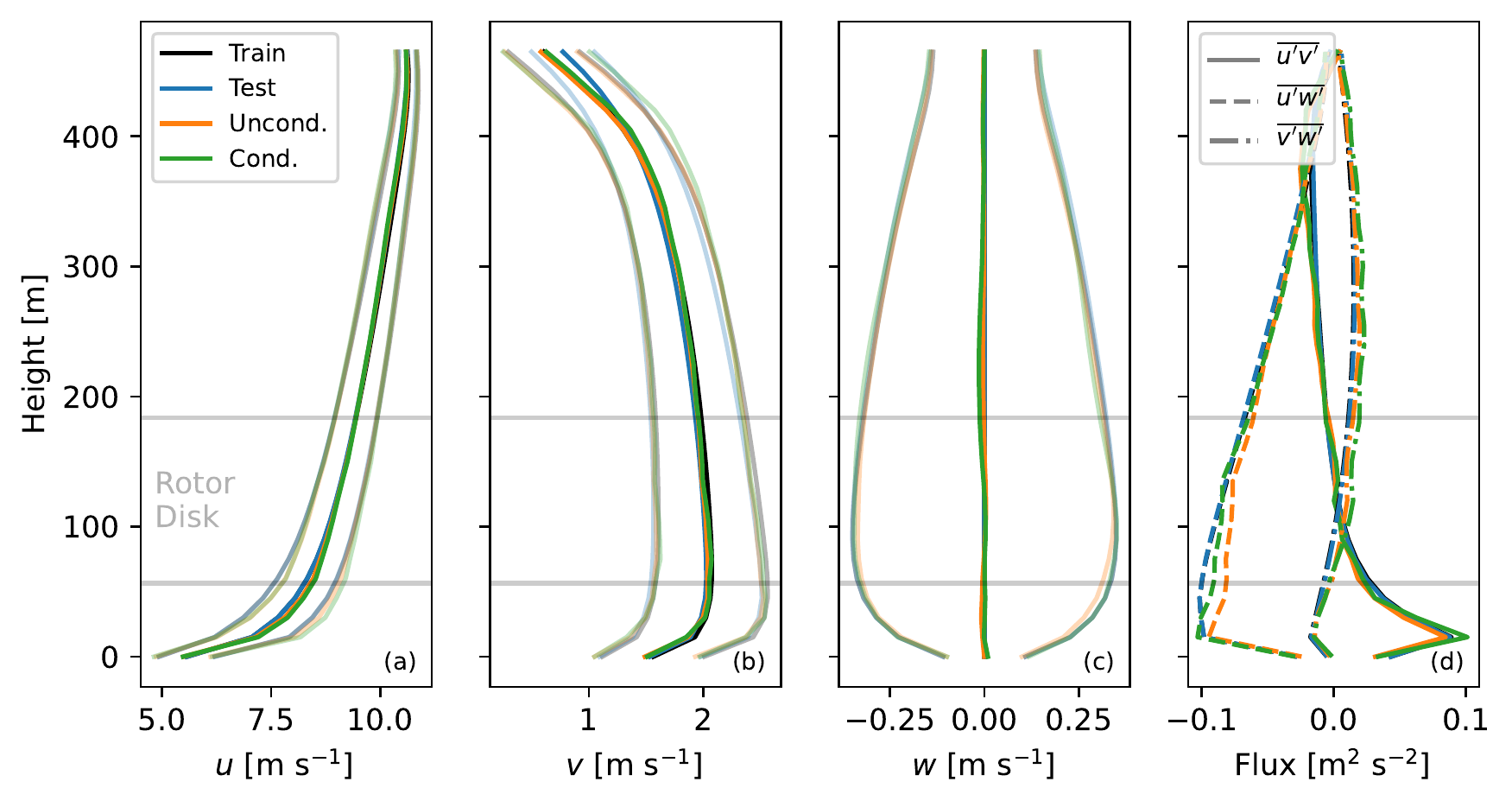}
\caption{(a)--(c) Vertical profiles of horizontally averaged velocity components from training data, testing data, unconditional LDM samples, and FC-mask conditional LDM samples. Averages are shown as solid lines, and $\pm$1 standard deviations are shown as faded lines. (d) Kinematic fluxes of the same data.}
\label{fig:les_profiles}
\end{figure*}

The LDM accurately captures average velocity profiles (Fig. \ref{fig:les_profiles}a--c), while slightly underperforming on kinematic velocity profiles (Fig. \ref{fig:les_profiles}d). We compare horizontally averaged profiles from the training dataset, testing dataset, $n=100$ unconditional samples, and $n=100$ conditional samples from the FC mask. The mean and standard deviation of LDM velocity profiles agree well with training data at all heights, and with testing data beneath the capping inversion. Thus, the LDM performs well on first-order statistics. However, the LDMs do show some deviations when comparing kinematic fluxes, a second-order statistic that is more challenging to capture than the mean. In particular, the unconditional LDM underestimates the downward flux $\overline{u'w'}$ by as much as 20\% between heights of 15 m and 150 m. The conditional LDM performs better in this region with only an 8\% discrepancy, possibly because of the near-surface conditioning information.

\begin{figure*}[htbp]
\centering
\includegraphics[width=3.5in]{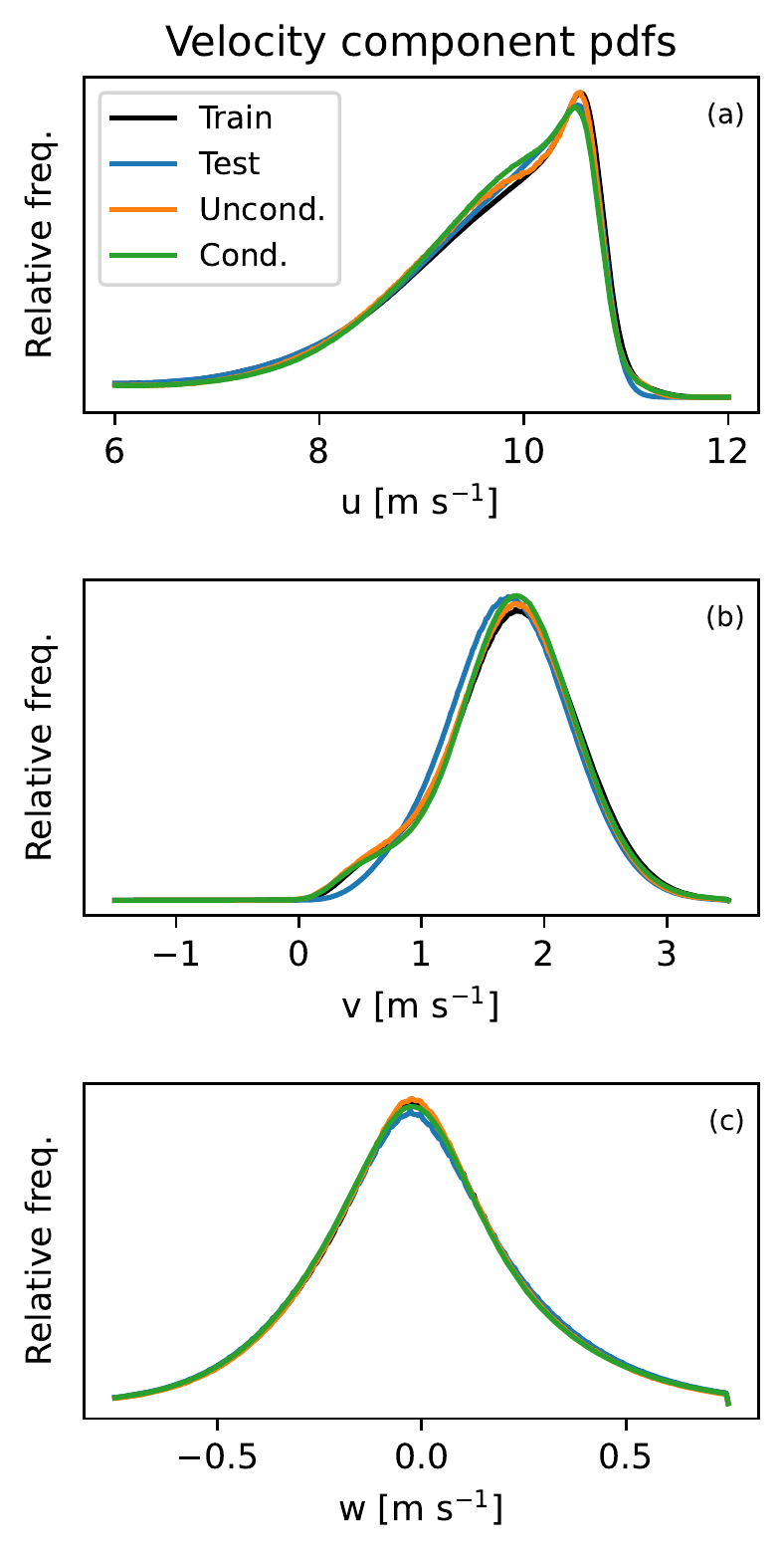}
\caption{The probability distribution function for training data, testing data, unconditional LDM samples, and conditional LDM samples for (a) $u$, (b) $v$, and (c) $w$.}
\label{fig:les_pdfs}
\end{figure*}

Similarly, the probability distribution functions (pdfs) of velocity components show broad agreement, while missing fine-scale details (Fig. \ref{fig:les_pdfs}). We calculate velocity distributions using the same three-dimensional data that was used to calculate average vertical profiles. The pdfs from LDM samples match the pdfs from training data well in all bins, only showing minor discrepancies.

\begin{figure*}[htbp]
\centering
\includegraphics[width=3.5in]{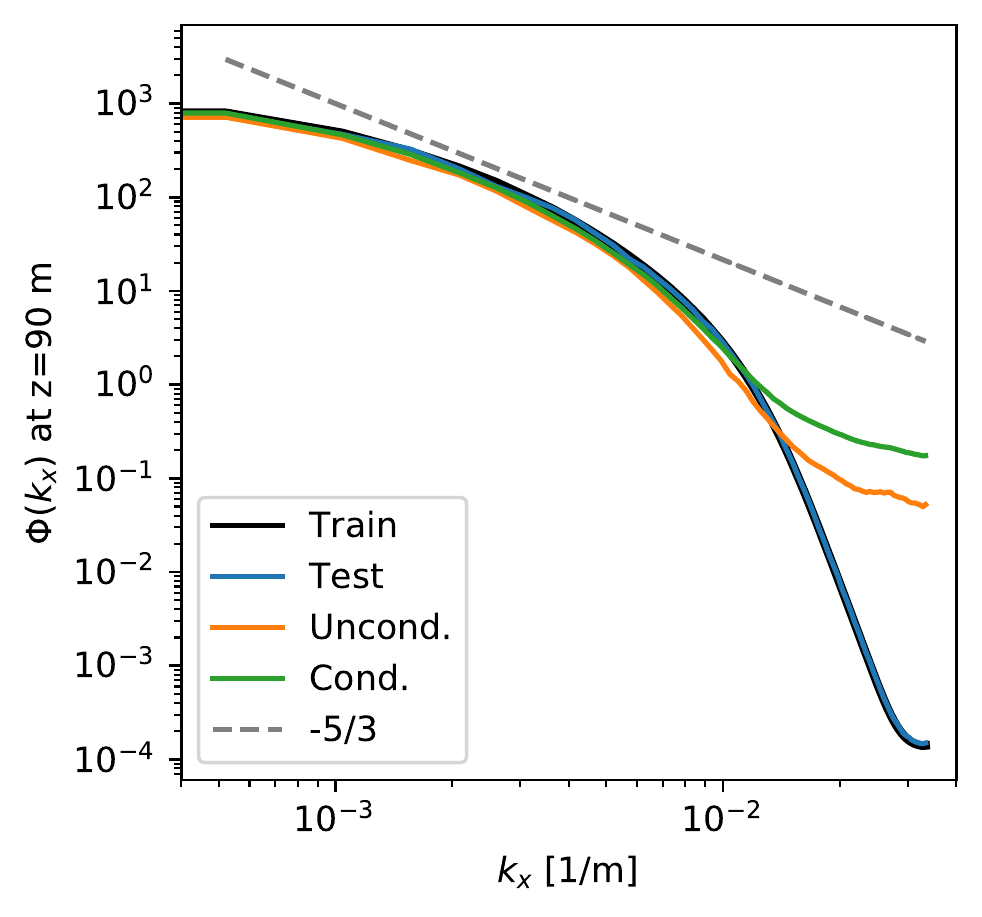}
\caption{The turbulence spectra of $u$ at $z=90$ m for training data, testing data, unconditional LDM samples, and conditional LDM samples.}
\label{fig:les_spectra}
\end{figure*}

While the power spectra of LDM samples match ground truth at large spatial scales, the LDM spectra have too much energy at the smallest spatial scales (Fig. \ref{fig:les_spectra}). We calculate one-dimensional spectra of $u$ at a height of 90 m. The LDM and ground truth spectra have similar values for wavenumbers smaller than 0.01 m$^{-1}$, and the LDM accurately captures the inertial scale of turbulence, as visualized by the $-5/3$ slope. However, for wavenumbers larger than 0.01 m$^{-1}$, the LDM has too much energy. This behavior has been observed in other machine learning reconstruction studies\cite{fukamiSuperresolutionReconstructionTurbulent2019,fukamiMachinelearningbasedSpatiotemporalSuper2021} and is consistent with the ``spectral bias'' problem \cite{rahamanSpectralBiasNeural2019} inherent to many neural networks---the autoencoder part of our network seeks to minimize the $L_1$ of the samples, thereby prioritizing reconstructing fields on the largest spatial scales. As such, the smallest scales have a relatively small impact on the loss function.

\begin{figure*}[htbp]
\centering
\includegraphics[width=3.5in]{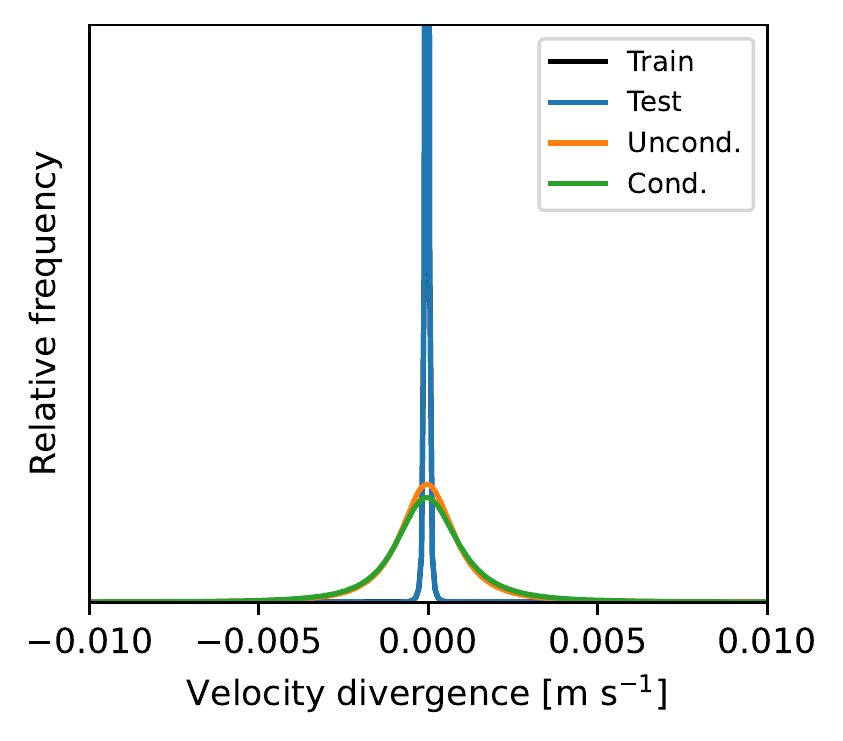}
\caption{An assessment of mass conservation, as visualized by distributions of velocity divergence for training data, testing data, unconditional LDM samples, and conditional LDM samples. Training and testing distributions sit on top of one another.}
\label{fig:les_mass}
\end{figure*}

Finally, we assess the LDM's ability to satisfy mass conservation by examining distributions of velocity divergence $\frac{\partial u_i}{\partial x_i}$ (Fig. \ref{fig:les_mass}). If continuity were exactly satisfied, the distribution of divergence would be shaped like a Dirichlet function at $\frac{\partial u_i}{\partial x_i} = 0$. The testing and training data approximately satisfy this behavior, showing small deviations from 0 due to numerical artifacts. While distributions for both the unconditional and conditional LDM data show a spike centered on 0, their spikes are not nearly as sharp. Thus, the LDMs appear to have some awareness of mass conservation, but they do not satisfy it here as well as the ground truth data.


\subsection{Assessment of ensembles}
\label{sec:ens_assessment}

\subsubsection{Rank histogram analysis}
For ensemble-based flow reconstruction, it is important to have a well-calibrated ensemble \cite{wilksStatisticalMethodsAtmospheric2019}. Ensemble members are drawn from some pdf, and in an ideal scenario, the ground truth would be seen as just another draw from that pdf, such that the ground truth is statistically indistinguishable from the ensemble members, a condition termed ``ensemble consistency.'' Ideally, the simulated ensemble would show zero bias relative to the ground truth, and the ensemble should have an appropriate but not excessive amount of spread. 

One common approach to assess ensemble consistency in the geosciences is the rank histogram \citep{hamillInterpretationRankHistograms2001,leinonenLatentDiffusionModels2023}. When given a group of predictions, the rank of the ground truth is calculated by sorting the ensemble members from smallest to largest by some scalar quantity (e.g., $u$ at a particular grid cell), and then identifying the position of the ground truth relative to the ensemble members. When repeated for several reconstructions, rank calculations can be collected into a rank histogram, a diagram that can diagnose ensemble bias as well as overconfidence or underconfidence.

\begin{figure*}[htbp]
\centering
\includegraphics[width=3.5in]{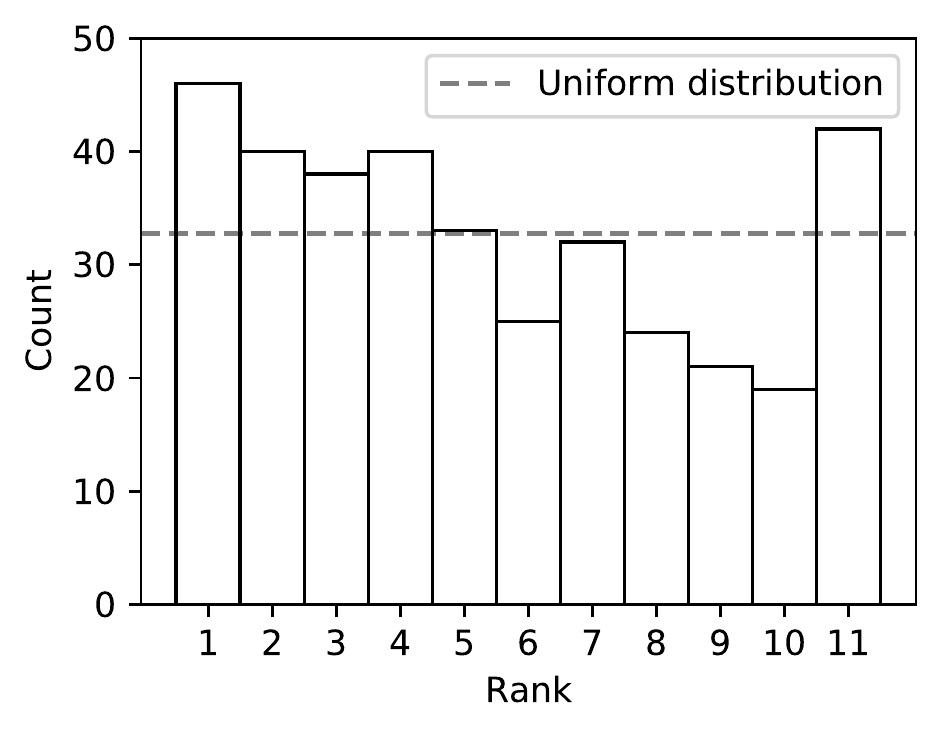}
\caption{Rank histogram for $u$ at the turbine nacelle, given 10-member LDM FC ensembles for 360 observations.}
\label{fig:rank_hist}
\end{figure*}

We assess ensemble quality by compiling a rank histogram using the first 6 hours (360 observations) from the test set and generating 10 ensemble members per observation (Fig. \ref{fig:rank_hist}). We calculate the rank of each ground truth by examining $u$ at the unobserved grid cell immediately downwind of the first scanning lidar observation, a cell that can be thought of as the turbine nacelle. An ideal rank histogram would look like a uniform distribution. However, the rank histogram here looks like a combination of a U-shape and a downward linear slope. The U-shape shows that LDM samples are underly diverse---the ground truth disproportionately falls at one extreme of the ensemble. The downward slope indicates that the ground truth tends to be disproportionately smaller than the rest of the ensemble, or in other words, the ensemble often overpredicts. We verify this by calculating the bias of the LDM samples relative to their respective ground truths, finding that $u$ is on average 0.02 m s$^{-1}$ higher. Thus, while the downward slope suggests a tendency to overpredict $u$ at turbine hub height, the near-zero bias is balanced out by the right arm of the U-shaped distribution. In the end, the LDM ensembles deviate from ideal behavior, which is in practice the case even for skillful ensemble forecasts \citep[e.g.,][]{monacheProbabilisticWeatherPrediction2013,ravuriSkilfulPrecipitationNowcasting2021}. These deviations may be small in the end, and future work will use these ensembles in a data assimilation process and assess if the unoptimal behavior is problematic in practice.

\subsubsection{External estimate of conditional standard deviation}

While rank histograms assess ensemble consistency and are used primarily in the ensemble weather forecasting field, we can alternatively assess the ensemble quality by examining ``sample diversity''. This concept is commonly examined in the deep generative modeling field, and when applied to computer vision problems, metrics such as Inception Score \citep{salimansImprovedTechniquesTraining2016} are commonly used. Here, we assess sample diversity by simply examining standard deviation, which is a valuable measure for physics-based problems. In Sec. \ref{sec:qual_analysis}, we assessed LDM sample diversity by taking a single observation, e.g., the one in Fig \ref{fig:qual_cond_raaw_planes}a, and calculating the standard deviation of $u$ across 100 conditionally generated samples (Fig \ref{fig:qual_cond_raaw_planes}e). This conditional standard deviation was calculated using samples from the LDM for a given observation. Alternatively, by applying tools such as fully connected neural networks or stochastic estimation \citep{hassanalyAdversarialSamplingUnknown2022} to our training dataset, it is possible to estimate what the conditional standard deviation should be for a given observation, without the need to generate any samples. This function has no awareness of the LDM or the samples it generates, and as such, it serves as an external check on the sample-calculated conditional standard deviation. 

Following \citet{hassanalyAdversarialSamplingUnknown2022}, we obtain an external estimate (Fig \ref{fig:qual_cond_raaw_planes}f, l) of the conditional standard deviation for the observation in Fig \ref{fig:qual_cond_raaw_planes}a, g by training a fully connected neural network. We provide more details on the external estimate methodology in Appendix C. From the side (Fig \ref{fig:qual_cond_raaw_planes}f), this external estimate agrees well with the LDM sample standard deviation (Fig \ref{fig:qual_cond_raaw_planes}e), but it fills in weaknesses that arise from calculating the sample standard deviation with a finite number of samples. From the side, the two standard deviations agree well in magnitude everywhere: the lowest grid cells, the observation network, and above the capping inversion. Both estimates also show larger variance just upwind of the meteorological mast and above the scanning lidar, a behavior that arises due to the presence of a coherent structure. When viewed from the top however (Fig \ref{fig:qual_cond_raaw_planes}l), the external estimate disagrees with the sample standard deviation. This discrepancy arises because the LDM here treats observations as a soft constraint, whereas the external estimate treats observations more like a hard constraint. As such, the external estimate shows a near-zero standard deviation in pixels where observation are available. From both the side and the top, the external estimate clearly shows that the sensing network has an impact on the flow reconstructions close to the observations, up to a maximum reach of roughly 150 m in any direction. By comparing the two standard deviations, we build confidence that the LDM doesn't show major deficiencies in terms of sample diversity. This result complements the rank histogram analysis, suggesting that indeed the LDM is only slightly underly diverse.

\subsubsection{Influence of FC observations on reconstruction accuracy}
\begin{figure*}[htbp]
\centering
\includegraphics[width=6in]{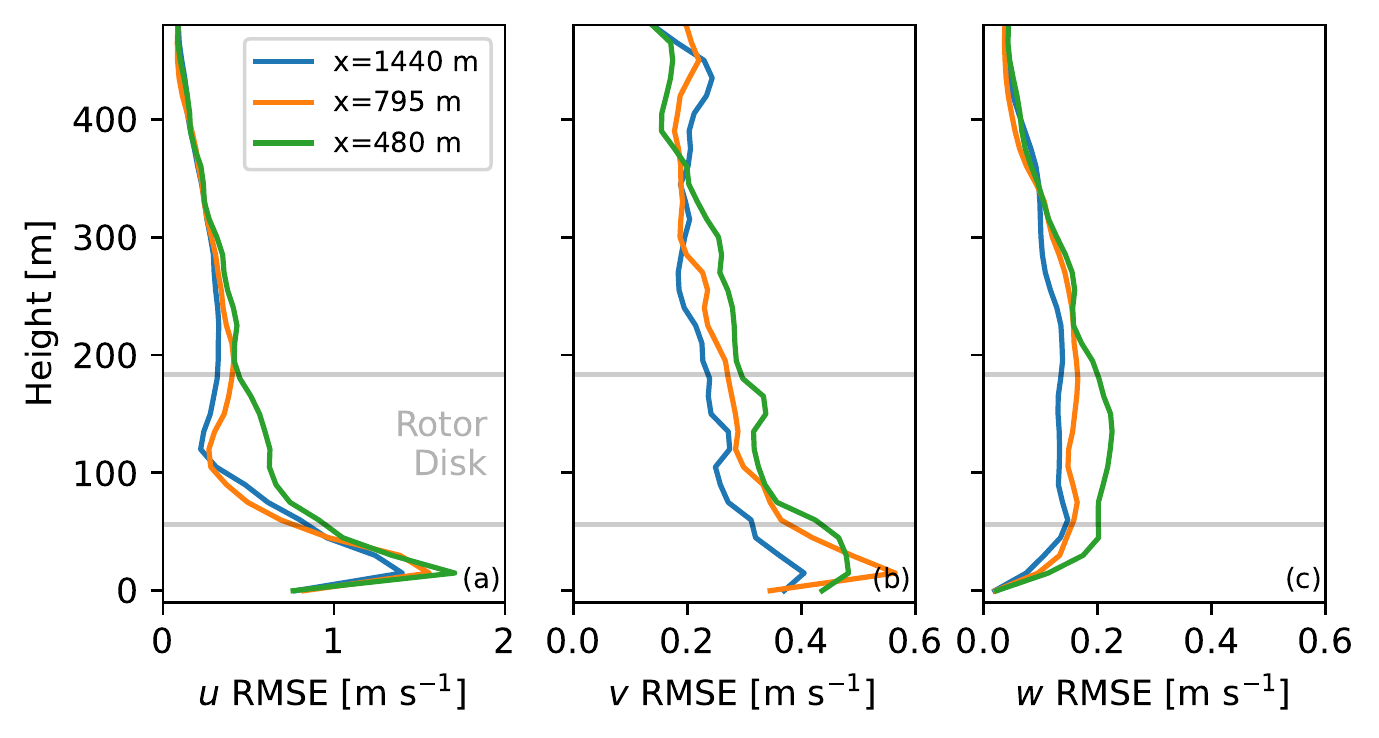}
\caption{Vertical profiles of root mean squared error for (a) $u$, (b) $v$, and (c) $w$ along the centerline of the domain at the meteorological mast ($x=1440$ m), the upstream edge of scanning lidar measurements ($x=795$ m), and a location upstream of all observations ($x=480$ m).}
\label{fig:rmse_profile}
\end{figure*}

Finally, we quantify the impact of FC observations on the reconstruction accuracy at three locations (Fig. \ref{fig:rmse_profile}). We calculate vertical profiles of root mean squared error (RMSE) using the dataset that was used in the rank histogram analysis: 360 ground truth samples from which observations are sampled, and 10 corresponding reconstructions for each ground truth. We calculate RMSE along the centerline of observations ($y=960$ m) at the location of the meteorological mast ($x=1440$ m), the upstream edge of scanning lidar measurements ($x=795$ m), and an upstream location where conditional standard deviation analysis revealed that observations no longer had an influence on reconstructions ($x=480$ m).

The RMSE profiles show that observations improve reconstruction accuracy predominantly as would be expected. Far away from observations at $x=480$ m, the RMSE values for all three velocity components are generally largest when compared to the other locations. The shape of these RMSE profiles correlates well with velocity standard deviations (Fig. \ref{fig:les_profiles})---heights with larger standard deviations are also associated with larger RMSE for the uncorrelated location. Within the perimeter of the synthetic field campaign at $x=795$ m and $x=1440$ m, observations have their strongest impacts at rotor disk heights, with relatively small impacts above and below the rotor disk. Relative to the RMSE outside of the measurement perimeter, the RMSE of $u$ is decreased most substantially (approximately 0.40 m s$^{-1}$) at the height of the scanning lidar measurements, as observations of $u$ are abundant at this elevation. While the field campaign measurements predominantly measure $u$, the observations still marginally impact RMSE profiles of $v$ and $w$ at rotor disk heights, decreasing $v$ RMSE by approximately 0.05--0.01 m s$^{-1}$ at most and $w$ RMSE by approximately 0.1 m s$^{-1}$. The meteorological mast at $x=1440$ m has the only observations of $v$ and $w$, and correspondingly, RMSE for these variables slightly outperforms RMSE at $x=795$ m.

\section{Conclusion}
\label{sec:conclusion}

In this paper, we investigate turbulent flow reconstruction in the context of a synthetic field campaign in the atmospheric boundary layer in which only spatially limited observations are available. We demonstrate that latent diffusion models (LDMs) create diverse, three-dimensional turbulent fields that are convincingly similar to true LES fields. These reconstructions match many physical characteristics of LES, such vertical profiles and the largest spatial scales of the spectra. However, LDMs struggle at the smallest spatial scales, diverging from LES spectra in this region and failing to preserve continuity.

Our LDM work extends machine learning turbulent flow reconstruction literature in several key manners. In contrast to many of the deterministic approaches, LDMs generate diverse samples, which is important for turbulent environments. Our LDM reconstructed samples with minimal observations ($<1\%$ of the domain is observed), though the algorithm also works well with abundant observations. And notably, LDM samples can be used as initial conditions for computationally stable LES simulations.  

This study suggests several lines of possible future inquiry for turbulent flow reconstruction. In upcoming work, we will explore applying LDMs to noisy, real-world measurements in the RAAW field campaign. The LDM architecture could be modified to improve performance at the smallest spatial scales, perhaps through the use of physics-informed losses\cite{raissiPhysicsinformedNeuralNetworks2019}. Finally, diffusion model researchers are investigating methods for quicker sampling \cite[e.g.,][]{mengDistillationGuidedDiffusion2022}, which could potentially open the road to real-time flow reconstruction. In summary, the results indicate that diffusion models are a powerful class of machine learning algorithm that is worthy of further exploration in turbulence research.


%
%

%

\begin{acknowledgments}
This work was authored in part by the National Renewable Energy Laboratory, operated by Alliance for Sustainable Energy, LLC, for the U.S. Department of Energy (DOE) under Contract No. DE-AC36-08GO28308. Funding provided by the U.S. Department of Energy Office of Energy Efficiency and Renewable Energy Wind Energy Technologies Office. The views expressed in the article do not necessarily represent the views of the DOE or the U.S. Government. The U.S. Government retains and the publisher, by accepting the article for publication, acknowledges that the U.S. Government retains a nonexclusive, paid-up, irrevocable, worldwide license to publish or reproduce the published form of this work, or allow others to do so, for U.S. Government purposes.

The research was performed using computational resources sponsored by the Department of Energy's Office of Energy Efficiency and Renewable Energy and located at the National Renewable Energy Laboratory. This work would not have been possible without several open-source diffusion modeling resources, including those by HuggingFace (https://huggingface.co/blog/annotated-diffusion), Janspiry (https://github.com/Janspiry/Palette-Image-to-Image-Diffusion-Models), Walter H. L. Pinaya (https://huggingface.co/spaces/Warvito/diffusion\_brain), and CompVis (https://github.com/CompVis/latent-diffusion). The analysis was performed using a number of open-source Python libraries, including numpy \citep{harrisArrayProgrammingNumPy2020}), Xarray \citep{hoyerXarrayNDLabeled2017}, and matplotlib \citep{hunterMatplotlib2DGraphics2007}. Thank you to Hristo Chipilski for helpful discussions regarding data assimilation, and Michael Kuhn for AMR-Wind support. Additionally, thank you to the editor and the reviewers in helping craft this manuscript.
\end{acknowledgments}

\appendix

\section{Additional visualizations of test data and unconditional samples}

\begin{figure*}[htbp]
\centering
\includegraphics[width=6.3in]{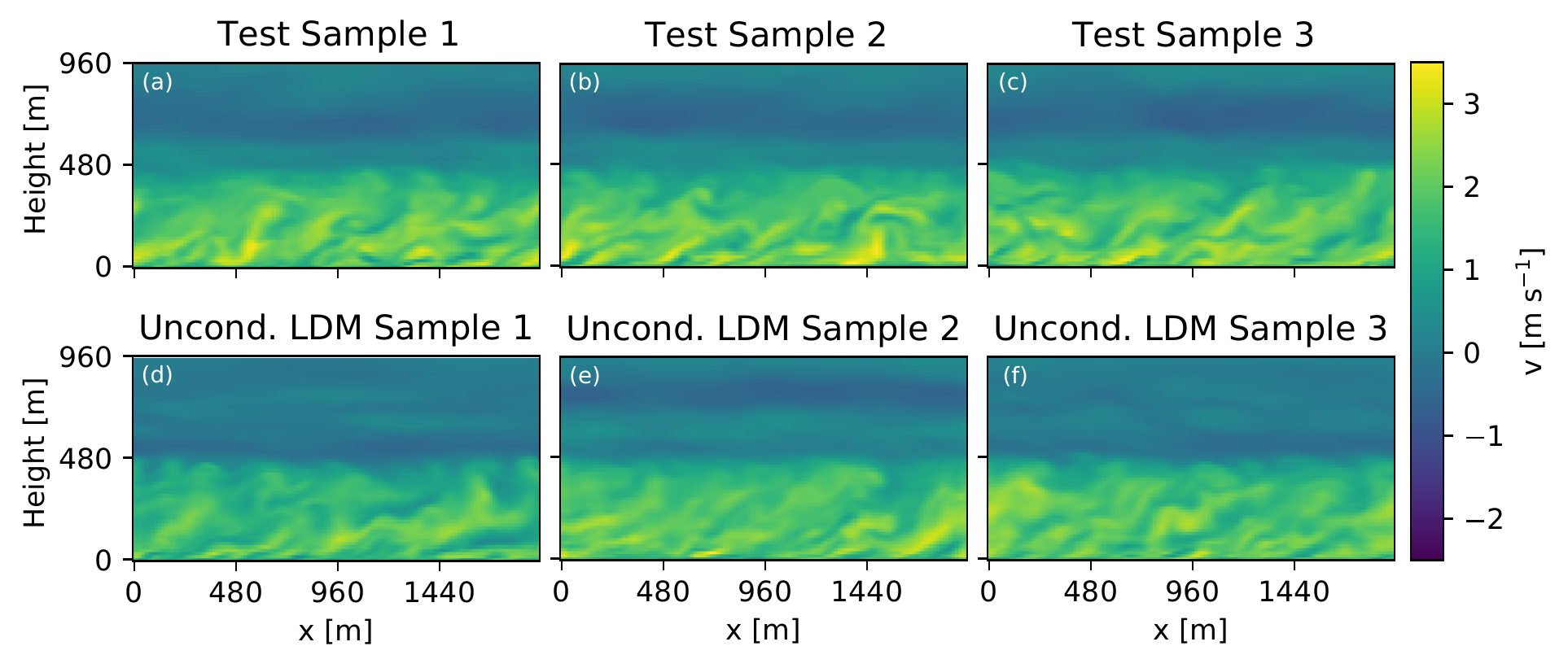}
\caption{(a)--(f) Same as Figure \ref{fig:qual_uncond_planes}, but for $v$.}
\label{fig:qual_cond_box_planes_v}
\end{figure*}

\begin{figure*}[htbp]
\centering
\includegraphics[width=6.3in]{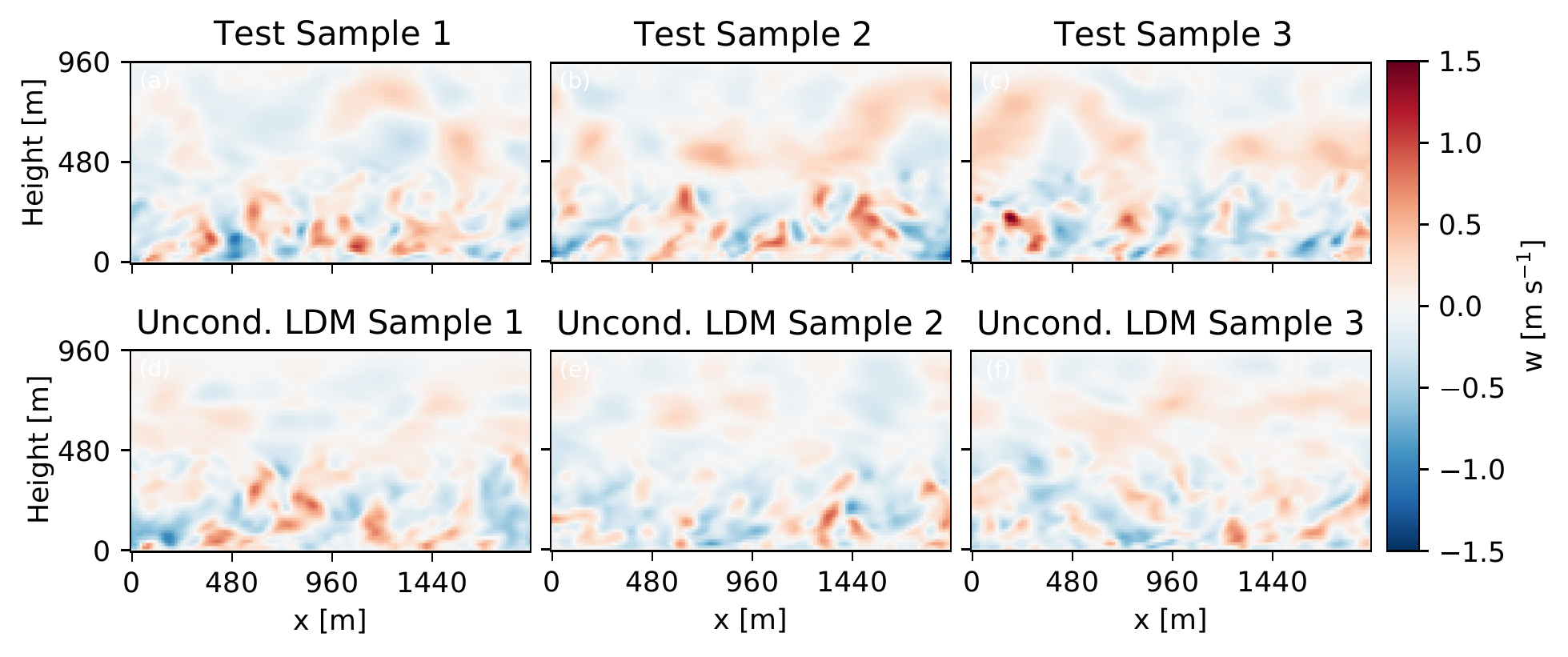}
\caption{(a)--(f) Same as Figure \ref{fig:qual_uncond_planes}, but for $w$.}
\label{fig:qual_cond_box_planes_w}
\end{figure*}

\begin{figure*}[htbp]
\centering
\includegraphics[width=6.3in]{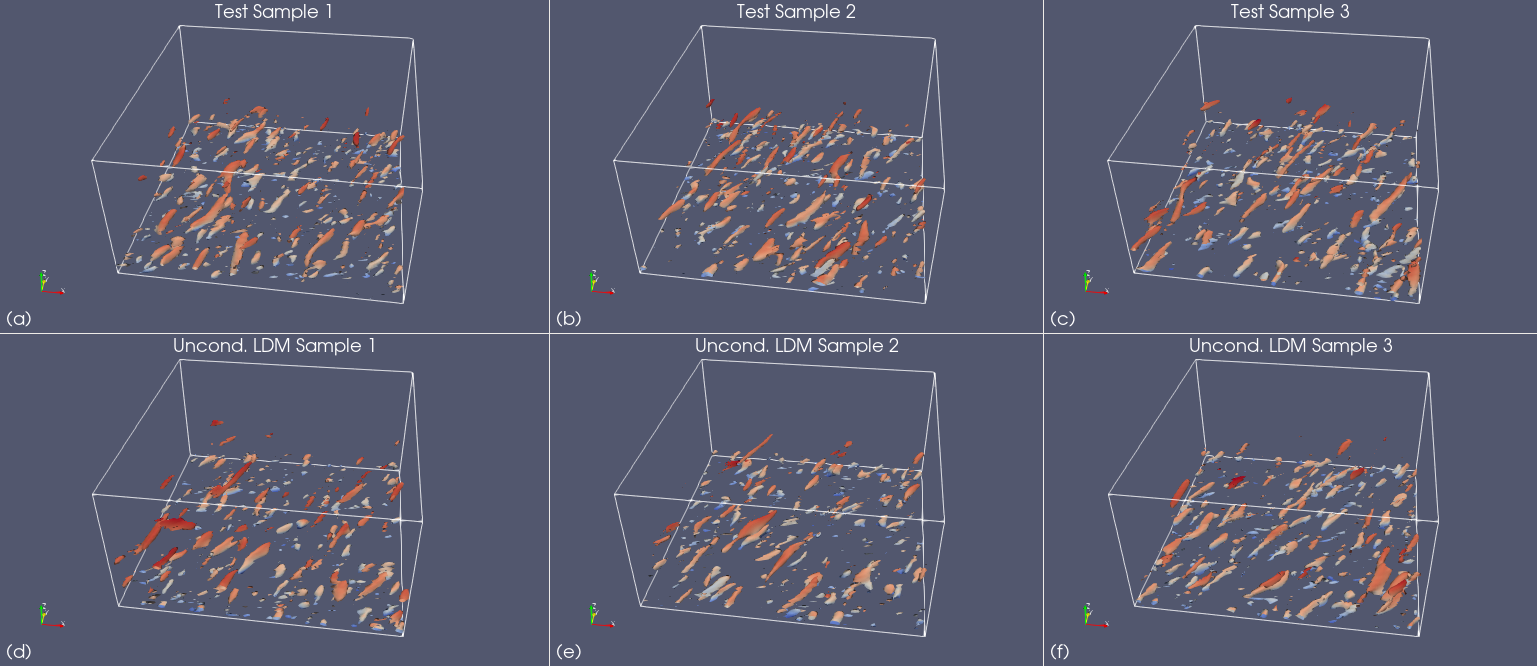}
\caption{(a)--(f) Isometric views of vortices identified as contours of the Q-criterion \citep{jeongIdentificationVortex1995} with a value of 0.0003 m s$^{-1}$.}
\label{fig:qual_iso}
\end{figure*}

\newpage\phantom{blabla}

\newpage\phantom{blabla}

\section{Additional visualizations of box and FC conditional samples}

\begin{figure*}[htbp]
\centering
\includegraphics[width=6.3in]{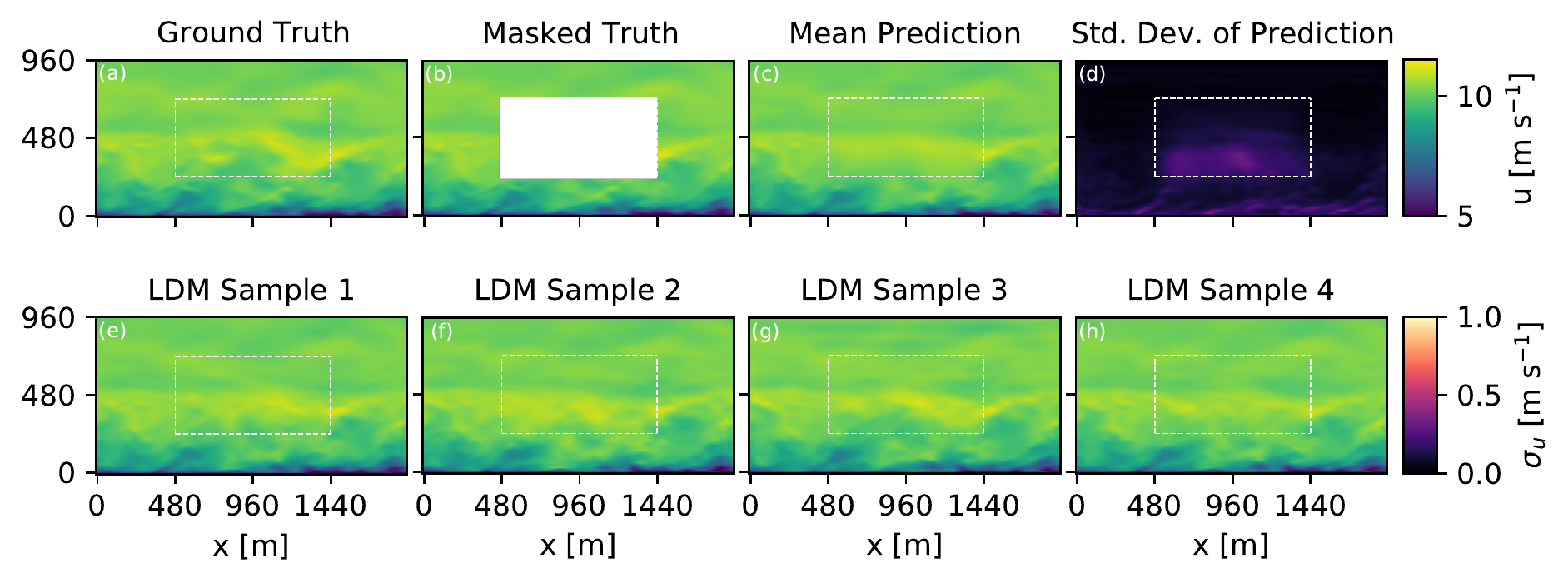}
\caption{A distinct (a) ground truth and (b) box observation from the one shown in Fig. \ref{fig:qual_cond_box_planes}. The (c) mean prediction and (d) standard deviation of prediction from 100 LDM samples are shown. (e)--(h) Four distinct LDM samples that are conditioned on the observation.}
\label{fig:qual_cond_box_planes2}
\end{figure*}

\begin{figure*}[htbp]
\centering
\includegraphics[width=6.3in]{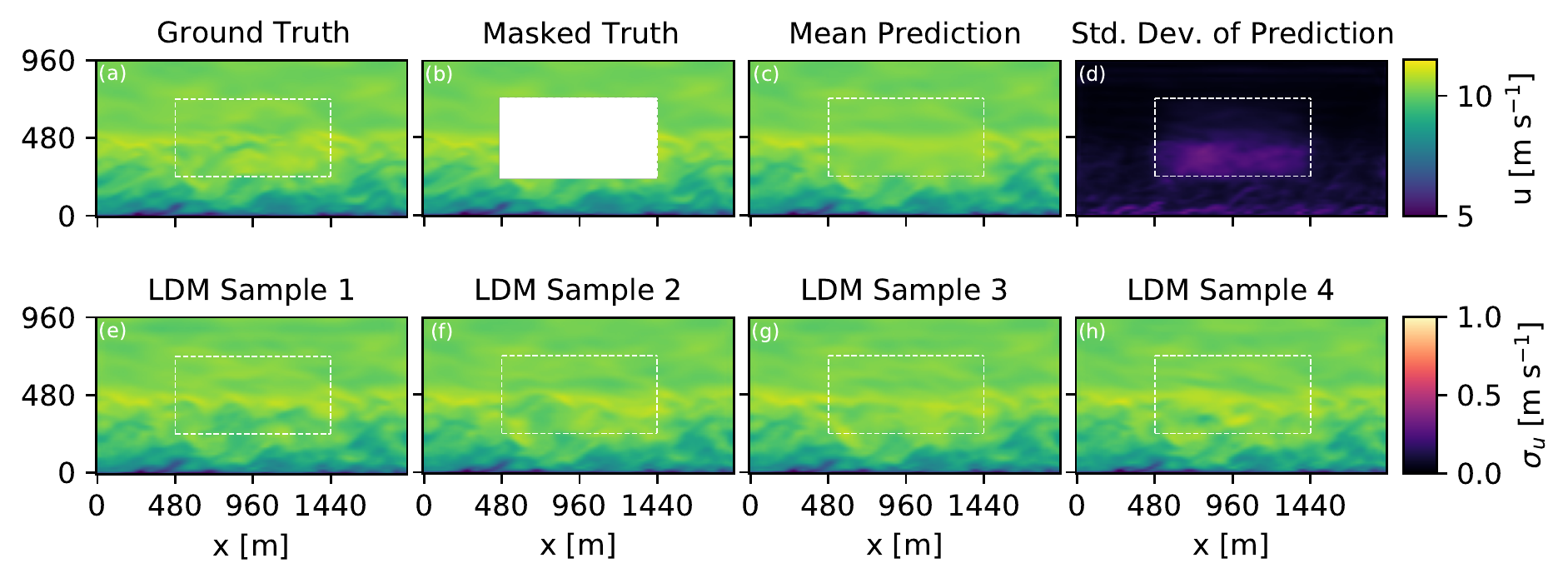}
\caption{(a)--(h) Same as Fig. \ref{fig:qual_cond_box_planes2}, except for a distinct observation.}
\end{figure*}

\begin{figure*}[htbp]
\centering
\includegraphics[width=6.3in]{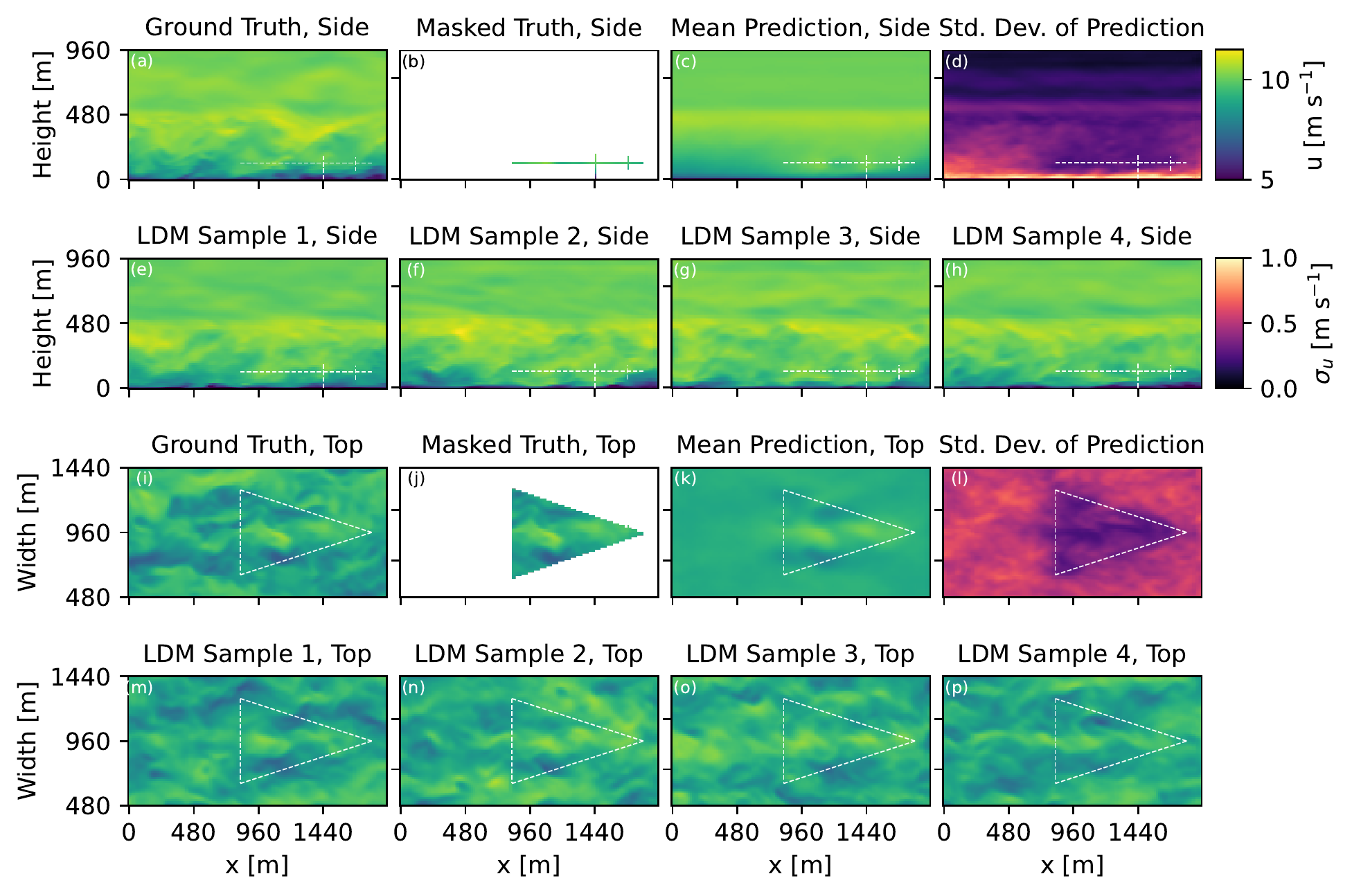}
\caption{A distinct (a) ground truth and (b) FC observation of $u$ from the one shown in Fig. \ref{fig:qual_cond_raaw_planes}. The (c) mean prediction and (d) standard deviation of prediction from 100 LDM samples are shown. (e)--(h) Four distinct LDM samples that are conditioned on the observation.}
\label{fig:qual_cond_raaw_planes2}
\end{figure*}

\begin{figure*}[htbp]
\centering
\includegraphics[width=6.3in]{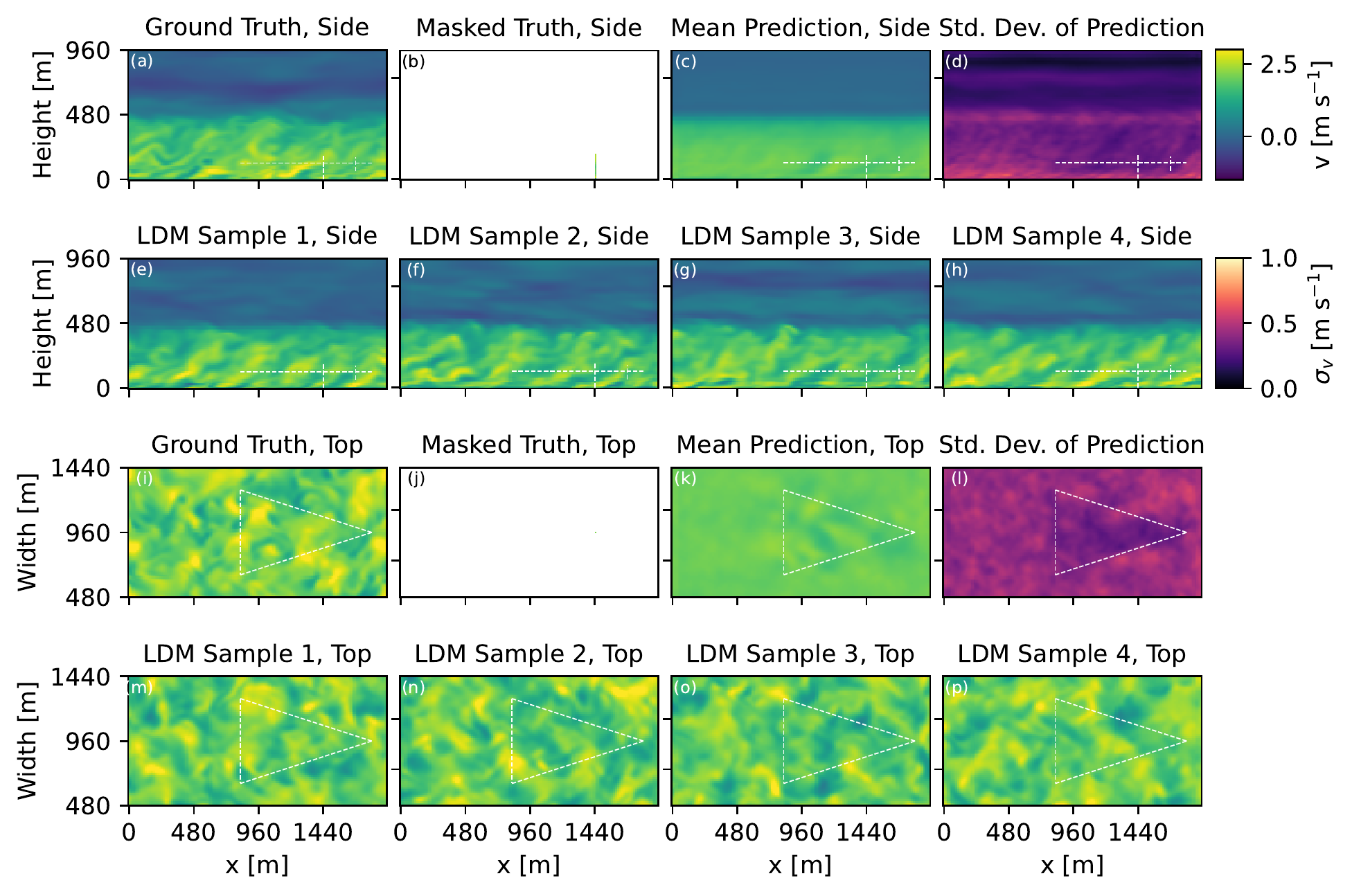}
\caption{(a)--(p) The same as Fig. \ref{fig:qual_cond_raaw_planes2}, except for $v$.}
\label{fig:qual_cond_raaw_planes2_v}
\end{figure*}

\begin{figure*}[htbp]
\centering
\includegraphics[width=6.3in]{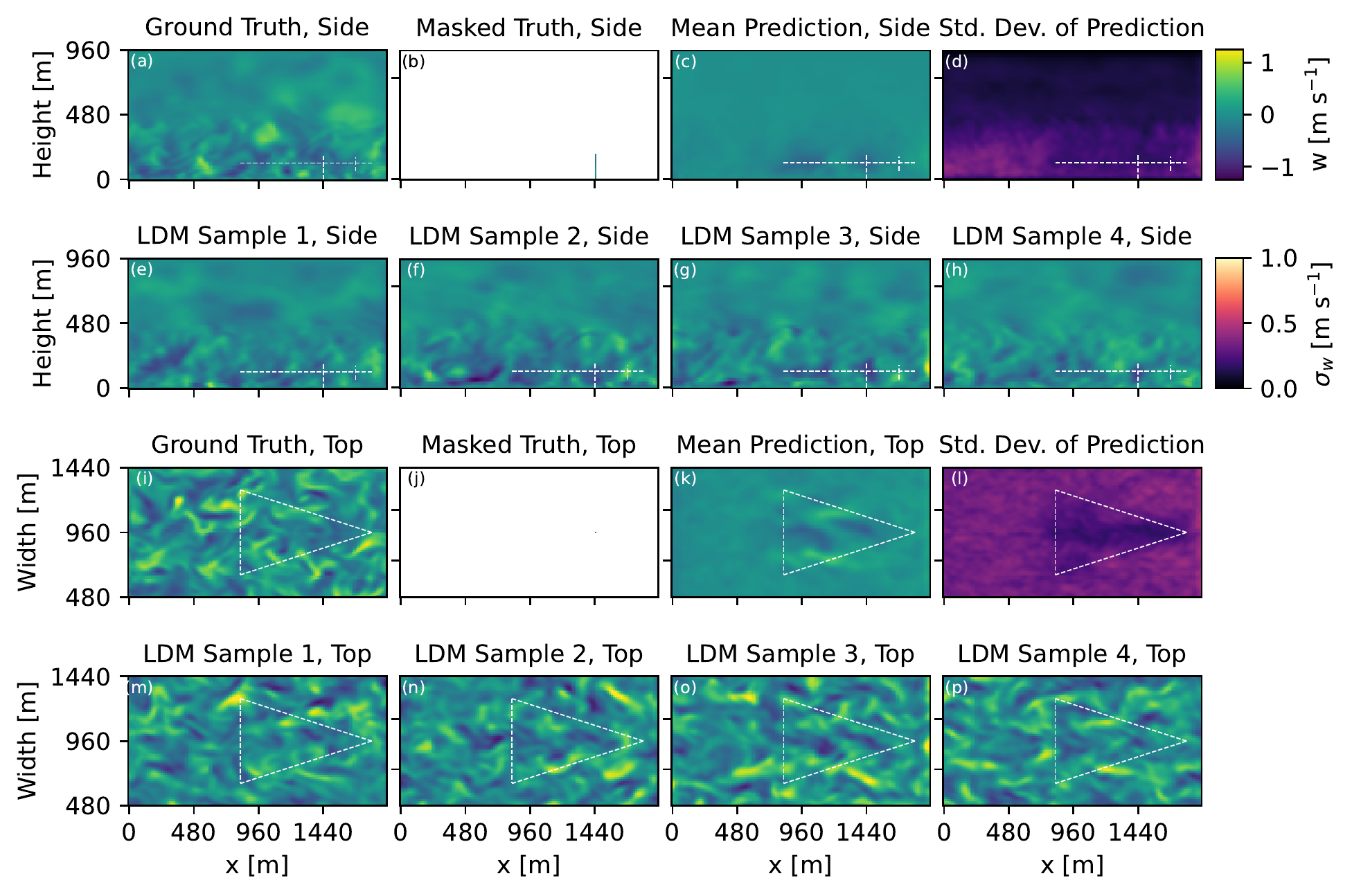}
\caption{(a)--(p) The same as Fig. \ref{fig:qual_cond_raaw_planes2}, except for $w$.}
\label{fig:qual_cond_raaw_planes2_w}
\end{figure*}


\newpage\phantom{blabla}

\newpage\phantom{blabla}

\newpage\phantom{blabla}

\newpage\phantom{blabla}

\section{Technical details for external estimates of the conditional standard deviation}
The a priori moment estimation method is detailed here. Consider an observation $y \in R^d$ where $d$ is the length of the observation vector, and $x \in R^m$ where $m$ is the length of the state vector. The function $f$ that best approximates $x$ given $y$, i.e., that minimizes $||f(y)-x||_2$, is called the optimal estimator of $x$ given $y$. In addition, the optimal estimator is the conditional expectation of $x$ given $y$ \citep{papoulis2002probability}. Therefore, by seeking the optimal estimator of $x$ given $y$, one can approximate the conditional moments of $x$ given $y$. Likewise, by seeking the conditional moments of $x^2$ given $y$, one can approximate the diagonal of the covariance matrix of the distribution sampled (i.e., the pointwise variance).

We obtain a priori estimates of conditional standard deviation by using a neural network. We construct the neural network to approximate the optimal estimator of the atmospheric state (for the first moment) and the square of the atmospheric state (for the second moment), given the observed field as the input. We use the architecture of \citet{Ledig_2017_CVPR}, a fully convolutional network with residual blocks and skip connections.

\begin{figure*}[htbp]
\centering
\includegraphics[width=0.9\textwidth]{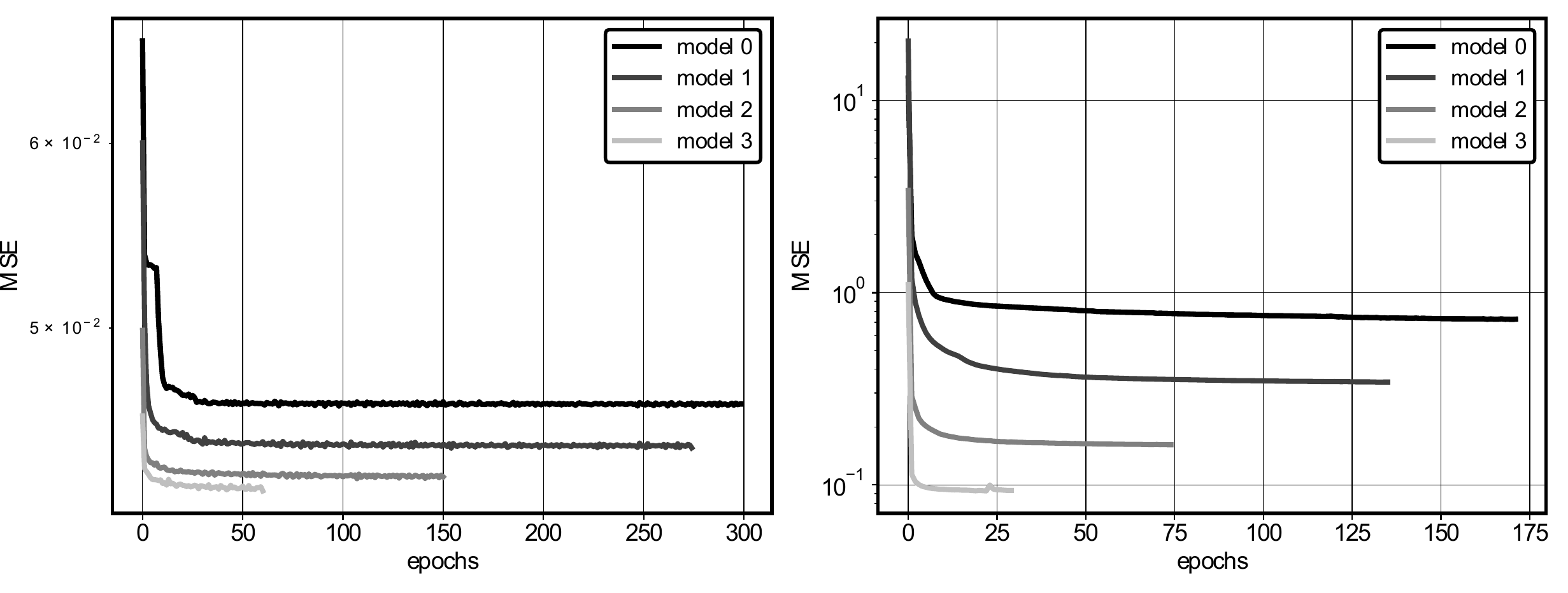}
\caption{MSE training loss versus epoch number for all the model architectures for FC mask for the first moment estimate (left) and the second moment estimate (right).}
\label{fig:trainingConv}
\end{figure*}

To ensure that a reasonable approximation of the optimal estimator has been reached, we train multiple neural nets with increasing complexity, until the mean-square-error (MSE) loss stops improving. Similar to \citet{hassanalyAdversarialSamplingUnknown2022}, we adjust two model architecture hyperparameters to converge to the optimal estimator: the number of filters per convolutional layers varied in the set $\{ 1,  2, 4, 8 \}$ and the number of residual blocks varied in the set $\{ 2, 4, 8, 16\}$. Unlike in \citet{hassanalyAdversarialSamplingUnknown2022}, the input and output of the network have the same dimension. The input is altered by setting the masked pixels to a zero value. The loss function is calculated by computing the MSE with the masked pixels only, thereby enforcing that unmasked pixels do not contribute to evaluating the conditional estimator. Since 3D convolutions are used and to accommodate memory requirements, the batch size is typically set to $8$ but changed to $4$ for the biggest network. A fixed learning rate of $1e-3$ was used. Similar to \citet{hassanalyAdversarialSamplingUnknown2022}, the second moment is estimated by outputting the square of the state substracted with the first moment. 
All networks considered here were sufficiently trained to reach convergence (Figure~\ref{fig:trainingConv}). 

\begin{figure*}[htbp]
\centering
\includegraphics[width=0.9\textwidth]{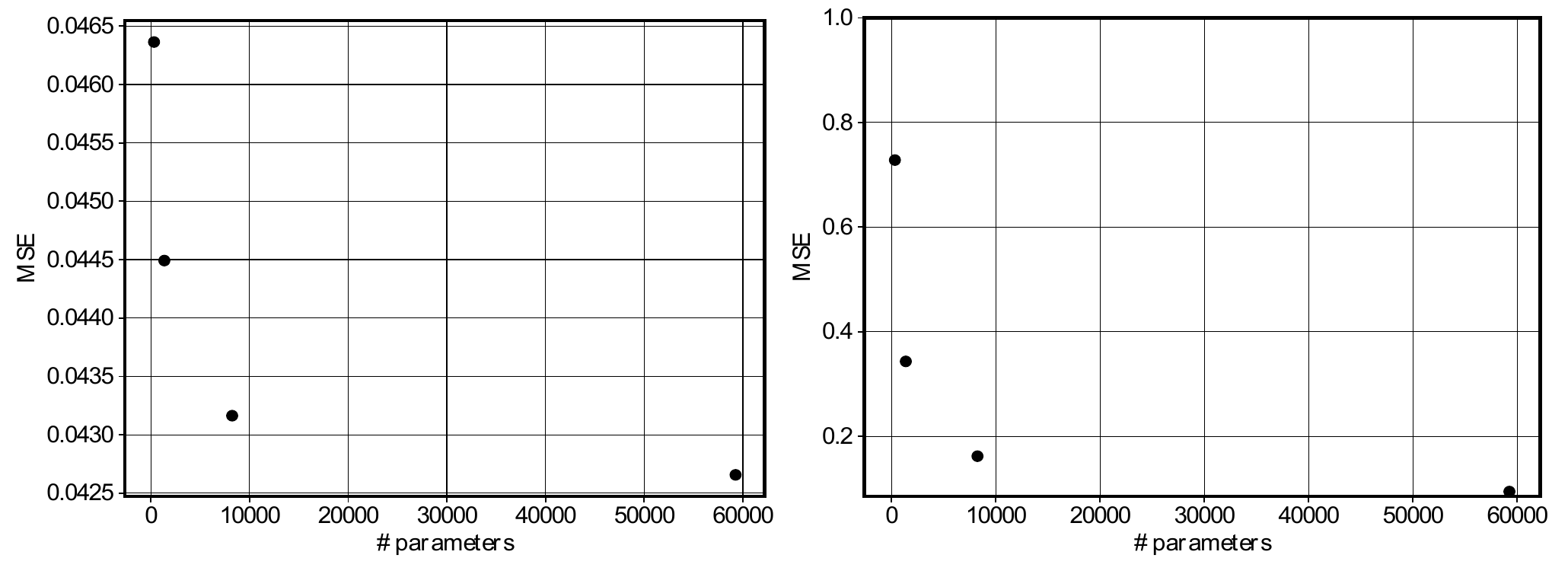}
\caption{Final MSE training loss versus number of trainable parameters for the FC mask for the first moment estimate (left) and the second moment estimate (right).}
\label{fig:convOptEst}
\end{figure*}

Once fully trained, the final training loss is examined to decide if further increase of network complexity is warranted to appropriately approximate the optimal estimator (Figure \ref{fig:convOptEst}). We find that as the number of trainable parameters increases beyond approximately 9,000 to approximately 60,000, the MSE decreases until a point where adding additional trainable parameters only marginally improves the loss. As such we stop increasing the complexity of the estimator network and produce the figure shown in Fig.~\ref{fig:qual_cond_raaw_planes}f using the model with the highest number of parameters.

\bibliography{mylibrary}

\begin{thebibliography}{68}%
\makeatletter
\providecommand \@ifxundefined [1]{%
 \@ifx{#1\undefined}
}%
\providecommand \@ifnum [1]{%
 \ifnum #1\expandafter \@firstoftwo
 \else \expandafter \@secondoftwo
 \fi
}%
\providecommand \@ifx [1]{%
 \ifx #1\expandafter \@firstoftwo
 \else \expandafter \@secondoftwo
 \fi
}%
\providecommand \natexlab [1]{#1}%
\providecommand \enquote  [1]{``#1''}%
\providecommand \bibnamefont  [1]{#1}%
\providecommand \bibfnamefont [1]{#1}%
\providecommand \citenamefont [1]{#1}%
\providecommand \href@noop [0]{\@secondoftwo}%
\providecommand \href [0]{\begingroup \@sanitize@url \@href}%
\providecommand \@href[1]{\@@startlink{#1}\@@href}%
\providecommand \@@href[1]{\endgroup#1\@@endlink}%
\providecommand \@sanitize@url [0]{\catcode `\\12\catcode `\$12\catcode `\&12\catcode `\#12\catcode `\^12\catcode `\_12\catcode `\%12\relax}%
\providecommand \@@startlink[1]{}%
\providecommand \@@endlink[0]{}%
\providecommand \url  [0]{\begingroup\@sanitize@url \@url }%
\providecommand \@url [1]{\endgroup\@href {#1}{\urlprefix }}%
\providecommand \urlprefix  [0]{URL }%
\providecommand \Eprint [0]{\href }%
\providecommand \doibase [0]{http://dx.doi.org/}%
\providecommand \selectlanguage [0]{\@gobble}%
\providecommand \bibinfo  [0]{\@secondoftwo}%
\providecommand \bibfield  [0]{\@secondoftwo}%
\providecommand \translation [1]{[#1]}%
\providecommand \BibitemOpen [0]{}%
\providecommand \bibitemStop [0]{}%
\providecommand \bibitemNoStop [0]{.\EOS\space}%
\providecommand \EOS [0]{\spacefactor3000\relax}%
\providecommand \BibitemShut  [1]{\csname bibitem#1\endcsname}%
\let\auto@bib@innerbib\@empty
\bibitem [{\citenamefont {Moriarty}\ \emph {et~al.}(2020)\citenamefont {Moriarty}, \citenamefont {Hamilton}, \citenamefont {Debnath}, \citenamefont {Herges}, \citenamefont {Isom}, \citenamefont {Lundquist}, \citenamefont {Maniaci}, \citenamefont {Naughton}, \citenamefont {Pauly}, \citenamefont {Roadman}, \citenamefont {Shaw}, \citenamefont {{van Dam}},\ and\ \citenamefont {Wharton}}]{moriartyAmericanWAKEExperimeNt2020}%
  \BibitemOpen
  \bibfield  {author} {\bibinfo {author} {\bibfnamefont {P.}~\bibnamefont {Moriarty}}, \bibinfo {author} {\bibfnamefont {N.}~\bibnamefont {Hamilton}}, \bibinfo {author} {\bibfnamefont {M.}~\bibnamefont {Debnath}}, \bibinfo {author} {\bibfnamefont {T.}~\bibnamefont {Herges}}, \bibinfo {author} {\bibfnamefont {B.}~\bibnamefont {Isom}}, \bibinfo {author} {\bibfnamefont {J.~K.}\ \bibnamefont {Lundquist}}, \bibinfo {author} {\bibfnamefont {D.}~\bibnamefont {Maniaci}}, \bibinfo {author} {\bibfnamefont {B.}~\bibnamefont {Naughton}}, \bibinfo {author} {\bibfnamefont {R.}~\bibnamefont {Pauly}}, \bibinfo {author} {\bibfnamefont {J.}~\bibnamefont {Roadman}}, \bibinfo {author} {\bibfnamefont {W.}~\bibnamefont {Shaw}}, \bibinfo {author} {\bibfnamefont {J.}~\bibnamefont {{van Dam}}}, \ and\ \bibinfo {author} {\bibfnamefont {S.}~\bibnamefont {Wharton}},\ }\href {\doibase 10.2172/1659798} {\enquote {\bibinfo {title} {American {{WAKE experimeNt}} ({{AWAKEN}})},}\ }\bibinfo {type} {Tech. Rep.}\ \bibinfo {number}
  {NREL/TP-5000-75789}\ (\bibinfo  {institution} {{National Renewable Energy Lab. (NREL), Golden, CO (United States)}},\ \bibinfo {year} {2020})\BibitemShut {NoStop}%
\bibitem [{\citenamefont {Hamilton}\ \emph {et~al.}(2022)\citenamefont {Hamilton}, \citenamefont {Doubrawa}, \citenamefont {Naughton},\ and\ \citenamefont {Kelley}}]{hamilton2022rotor}%
  \BibitemOpen
  \bibfield  {author} {\bibinfo {author} {\bibfnamefont {N.}~\bibnamefont {Hamilton}}, \bibinfo {author} {\bibfnamefont {P.}~\bibnamefont {Doubrawa}}, \bibinfo {author} {\bibfnamefont {J.}~\bibnamefont {Naughton}}, \ and\ \bibinfo {author} {\bibfnamefont {C.}~\bibnamefont {Kelley}},\ }\bibfield  {title} {\enquote {\bibinfo {title} {Rotor aerodynamics, aeroelastics, and wake ({{RAAW}}) campaign overview},}\ }\href@noop {} {\bibfield  {journal} {\bibinfo  {journal} {Bulletin of the American Physical Society}\ } (\bibinfo {year} {2022})}\BibitemShut {NoStop}%
\bibitem [{\citenamefont {Allwine}\ and\ \citenamefont {Flaherty}(2006)}]{allwineJointUrban20032006}%
  \BibitemOpen
  \bibfield  {author} {\bibinfo {author} {\bibfnamefont {K.~J.}\ \bibnamefont {Allwine}}\ and\ \bibinfo {author} {\bibfnamefont {J.~E.}\ \bibnamefont {Flaherty}},\ }\href {\doibase 10.2172/890732} {\enquote {\bibinfo {title} {Joint {{Urban}} 2003: {{Study Overview And Instrument Locations}}},}\ }\bibinfo {type} {Tech. Rep.}\ \bibinfo {number} {PNNL-15967}\ (\bibinfo  {institution} {{Pacific Northwest National Lab. (PNNL), Richland, WA (United States)}},\ \bibinfo {year} {2006})\BibitemShut {NoStop}%
\bibitem [{\citenamefont {Warneke}\ \emph {et~al.}(2023)\citenamefont {Warneke}, \citenamefont {Schwarz}, \citenamefont {Dibb}, \citenamefont {Kalashnikova}, \citenamefont {Frost}, \citenamefont {{Al-Saad}}, \citenamefont {Brown}, \citenamefont {Brewer}, \citenamefont {Soja}, \citenamefont {Seidel}, \citenamefont {Washenfelder}, \citenamefont {Wiggins}, \citenamefont {Moore}, \citenamefont {Anderson}, \citenamefont {Jordan}, \citenamefont {Yacovitch}, \citenamefont {Herndon}, \citenamefont {Liu}, \citenamefont {Kuwayama}, \citenamefont {Jaffe}, \citenamefont {Johnston}, \citenamefont {Selimovic}, \citenamefont {Yokelson}, \citenamefont {Giles}, \citenamefont {Holben}, \citenamefont {Goloub}, \citenamefont {Popovici}, \citenamefont {Trainer}, \citenamefont {Kumar}, \citenamefont {Pierce}, \citenamefont {Fahey}, \citenamefont {Roberts}, \citenamefont {Gargulinski}, \citenamefont {Peterson}, \citenamefont {Ye}, \citenamefont {Thapa}, \citenamefont {Saide}, \citenamefont {Fite}, \citenamefont {Holmes},
  \citenamefont {Wang}, \citenamefont {Coggon}, \citenamefont {Decker}, \citenamefont {Stockwell}, \citenamefont {Xu}, \citenamefont {Gkatzelis}, \citenamefont {Aikin}, \citenamefont {Lefer}, \citenamefont {Kaspari}, \citenamefont {Griffin}, \citenamefont {Zeng}, \citenamefont {Weber}, \citenamefont {Hastings}, \citenamefont {Chai}, \citenamefont {Wolfe}, \citenamefont {Hanisco}, \citenamefont {Liao}, \citenamefont {Campuzano~Jost}, \citenamefont {Guo}, \citenamefont {Jimenez}, \citenamefont {Crawford},\ and\ \citenamefont {Team}}]{warnekeFireInfluenceRegional2023}%
  \BibitemOpen
  \bibfield  {author} {\bibinfo {author} {\bibfnamefont {C.}~\bibnamefont {Warneke}}, \bibinfo {author} {\bibfnamefont {J.~P.}\ \bibnamefont {Schwarz}}, \bibinfo {author} {\bibfnamefont {J.}~\bibnamefont {Dibb}}, \bibinfo {author} {\bibfnamefont {O.}~\bibnamefont {Kalashnikova}}, \bibinfo {author} {\bibfnamefont {G.}~\bibnamefont {Frost}}, \bibinfo {author} {\bibfnamefont {J.}~\bibnamefont {{Al-Saad}}}, \bibinfo {author} {\bibfnamefont {S.~S.}\ \bibnamefont {Brown}}, \bibinfo {author} {\bibfnamefont {W.~A.}\ \bibnamefont {Brewer}}, \bibinfo {author} {\bibfnamefont {A.}~\bibnamefont {Soja}}, \bibinfo {author} {\bibfnamefont {F.~C.}\ \bibnamefont {Seidel}}, \bibinfo {author} {\bibfnamefont {R.~A.}\ \bibnamefont {Washenfelder}}, \bibinfo {author} {\bibfnamefont {E.~B.}\ \bibnamefont {Wiggins}}, \bibinfo {author} {\bibfnamefont {R.~H.}\ \bibnamefont {Moore}}, \bibinfo {author} {\bibfnamefont {B.~E.}\ \bibnamefont {Anderson}}, \bibinfo {author} {\bibfnamefont {C.}~\bibnamefont {Jordan}}, \bibinfo {author}
  {\bibfnamefont {T.~I.}\ \bibnamefont {Yacovitch}}, \bibinfo {author} {\bibfnamefont {S.~C.}\ \bibnamefont {Herndon}}, \bibinfo {author} {\bibfnamefont {S.}~\bibnamefont {Liu}}, \bibinfo {author} {\bibfnamefont {T.}~\bibnamefont {Kuwayama}}, \bibinfo {author} {\bibfnamefont {D.}~\bibnamefont {Jaffe}}, \bibinfo {author} {\bibfnamefont {N.}~\bibnamefont {Johnston}}, \bibinfo {author} {\bibfnamefont {V.}~\bibnamefont {Selimovic}}, \bibinfo {author} {\bibfnamefont {R.}~\bibnamefont {Yokelson}}, \bibinfo {author} {\bibfnamefont {D.~M.}\ \bibnamefont {Giles}}, \bibinfo {author} {\bibfnamefont {B.~N.}\ \bibnamefont {Holben}}, \bibinfo {author} {\bibfnamefont {P.}~\bibnamefont {Goloub}}, \bibinfo {author} {\bibfnamefont {I.}~\bibnamefont {Popovici}}, \bibinfo {author} {\bibfnamefont {M.}~\bibnamefont {Trainer}}, \bibinfo {author} {\bibfnamefont {A.}~\bibnamefont {Kumar}}, \bibinfo {author} {\bibfnamefont {R.~B.}\ \bibnamefont {Pierce}}, \bibinfo {author} {\bibfnamefont {D.}~\bibnamefont {Fahey}}, \bibinfo {author}
  {\bibfnamefont {J.}~\bibnamefont {Roberts}}, \bibinfo {author} {\bibfnamefont {E.~M.}\ \bibnamefont {Gargulinski}}, \bibinfo {author} {\bibfnamefont {D.~A.}\ \bibnamefont {Peterson}}, \bibinfo {author} {\bibfnamefont {X.}~\bibnamefont {Ye}}, \bibinfo {author} {\bibfnamefont {L.~H.}\ \bibnamefont {Thapa}}, \bibinfo {author} {\bibfnamefont {P.~E.}\ \bibnamefont {Saide}}, \bibinfo {author} {\bibfnamefont {C.~H.}\ \bibnamefont {Fite}}, \bibinfo {author} {\bibfnamefont {C.~D.}\ \bibnamefont {Holmes}}, \bibinfo {author} {\bibfnamefont {S.}~\bibnamefont {Wang}}, \bibinfo {author} {\bibfnamefont {M.~M.}\ \bibnamefont {Coggon}}, \bibinfo {author} {\bibfnamefont {Z.~C.~J.}\ \bibnamefont {Decker}}, \bibinfo {author} {\bibfnamefont {C.~E.}\ \bibnamefont {Stockwell}}, \bibinfo {author} {\bibfnamefont {L.}~\bibnamefont {Xu}}, \bibinfo {author} {\bibfnamefont {G.}~\bibnamefont {Gkatzelis}}, \bibinfo {author} {\bibfnamefont {K.}~\bibnamefont {Aikin}}, \bibinfo {author} {\bibfnamefont {B.}~\bibnamefont {Lefer}}, \bibinfo
  {author} {\bibfnamefont {J.}~\bibnamefont {Kaspari}}, \bibinfo {author} {\bibfnamefont {D.}~\bibnamefont {Griffin}}, \bibinfo {author} {\bibfnamefont {L.}~\bibnamefont {Zeng}}, \bibinfo {author} {\bibfnamefont {R.}~\bibnamefont {Weber}}, \bibinfo {author} {\bibfnamefont {M.}~\bibnamefont {Hastings}}, \bibinfo {author} {\bibfnamefont {J.}~\bibnamefont {Chai}}, \bibinfo {author} {\bibfnamefont {G.~M.}\ \bibnamefont {Wolfe}}, \bibinfo {author} {\bibfnamefont {T.~F.}\ \bibnamefont {Hanisco}}, \bibinfo {author} {\bibfnamefont {J.}~\bibnamefont {Liao}}, \bibinfo {author} {\bibfnamefont {P.}~\bibnamefont {Campuzano~Jost}}, \bibinfo {author} {\bibfnamefont {H.}~\bibnamefont {Guo}}, \bibinfo {author} {\bibfnamefont {J.~L.}\ \bibnamefont {Jimenez}}, \bibinfo {author} {\bibfnamefont {J.}~\bibnamefont {Crawford}}, \ and\ \bibinfo {author} {\bibfnamefont {T.~F.-A.~S.}\ \bibnamefont {Team}},\ }\bibfield  {title} {\enquote {\bibinfo {title} {Fire {{Influence}} on {{Regional}} to {{Global Environments}} and {{Air Quality}}
  ({{FIREX-AQ}})},}\ }\href {\doibase 10.1029/2022JD037758} {\bibfield  {journal} {\bibinfo  {journal} {Journal of Geophysical Research: Atmospheres}\ }\textbf {\bibinfo {volume} {128}},\ \bibinfo {pages} {e2022JD037758} (\bibinfo {year} {2023})}\BibitemShut {NoStop}%
\bibitem [{\citenamefont {Buzzicotti}(2023)}]{buzzicottiDataReconstructionComplex2023}%
  \BibitemOpen
  \bibfield  {author} {\bibinfo {author} {\bibfnamefont {M.}~\bibnamefont {Buzzicotti}},\ }\bibfield  {title} {\enquote {\bibinfo {title} {Data reconstruction for complex flows using {{AI}}: {{Recent}} progress, obstacles, and perspectives},}\ }\href {\doibase 10.1209/0295-5075/acc88c} {\bibfield  {journal} {\bibinfo  {journal} {Europhysics Letters}\ }\textbf {\bibinfo {volume} {142}},\ \bibinfo {pages} {23001} (\bibinfo {year} {2023})}\BibitemShut {NoStop}%
\bibitem [{\citenamefont {Yu}, \citenamefont {Bi},\ and\ \citenamefont {Fan}(2023)}]{yuDeepLearningFluid2023}%
  \BibitemOpen
  \bibfield  {author} {\bibinfo {author} {\bibfnamefont {C.}~\bibnamefont {Yu}}, \bibinfo {author} {\bibfnamefont {X.}~\bibnamefont {Bi}}, \ and\ \bibinfo {author} {\bibfnamefont {Y.}~\bibnamefont {Fan}},\ }\bibfield  {title} {\enquote {\bibinfo {title} {Deep learning for fluid velocity field estimation: {{A}} review},}\ }\href {\doibase 10.1016/j.oceaneng.2023.113693} {\bibfield  {journal} {\bibinfo  {journal} {Ocean Engineering}\ }\textbf {\bibinfo {volume} {271}},\ \bibinfo {pages} {113693} (\bibinfo {year} {2023})}\BibitemShut {NoStop}%
\bibitem [{\citenamefont {Meldi}\ and\ \citenamefont {Poux}(2017)}]{meldiReducedOrderModel2017}%
  \BibitemOpen
  \bibfield  {author} {\bibinfo {author} {\bibfnamefont {M.}~\bibnamefont {Meldi}}\ and\ \bibinfo {author} {\bibfnamefont {A.}~\bibnamefont {Poux}},\ }\bibfield  {title} {\enquote {\bibinfo {title} {A reduced order model based on {{Kalman}} filtering for sequential data assimilation of turbulent flows},}\ }\href {\doibase 10.1016/j.jcp.2017.06.042} {\bibfield  {journal} {\bibinfo  {journal} {Journal of Computational Physics}\ }\textbf {\bibinfo {volume} {347}},\ \bibinfo {pages} {207--234} (\bibinfo {year} {2017})}\BibitemShut {NoStop}%
\bibitem [{\citenamefont {Chandramouli}, \citenamefont {Memin},\ and\ \citenamefont {Heitz}(2020)}]{chandramouli4DLargeScale2020}%
  \BibitemOpen
  \bibfield  {author} {\bibinfo {author} {\bibfnamefont {P.}~\bibnamefont {Chandramouli}}, \bibinfo {author} {\bibfnamefont {E.}~\bibnamefont {Memin}}, \ and\ \bibinfo {author} {\bibfnamefont {D.}~\bibnamefont {Heitz}},\ }\bibfield  {title} {\enquote {\bibinfo {title} {{{4D}} large scale variational data assimilation of a turbulent flow with a dynamics error model},}\ }\href {\doibase 10.1016/j.jcp.2020.109446} {\bibfield  {journal} {\bibinfo  {journal} {Journal of Computational Physics}\ }\textbf {\bibinfo {volume} {412}},\ \bibinfo {pages} {109446} (\bibinfo {year} {2020})}\BibitemShut {NoStop}%
\bibitem [{\citenamefont {Mons}, \citenamefont {Du},\ and\ \citenamefont {Zaki}(2021)}]{monsEnsemblevariationalAssimilationStatistical2021}%
  \BibitemOpen
  \bibfield  {author} {\bibinfo {author} {\bibfnamefont {V.}~\bibnamefont {Mons}}, \bibinfo {author} {\bibfnamefont {Y.}~\bibnamefont {Du}}, \ and\ \bibinfo {author} {\bibfnamefont {T.~A.}\ \bibnamefont {Zaki}},\ }\bibfield  {title} {\enquote {\bibinfo {title} {Ensemble-variational assimilation of statistical data in large-eddy simulation},}\ }\href {\doibase 10.1103/PhysRevFluids.6.104607} {\bibfield  {journal} {\bibinfo  {journal} {Physical Review Fluids}\ }\textbf {\bibinfo {volume} {6}},\ \bibinfo {pages} {104607} (\bibinfo {year} {2021})}\BibitemShut {NoStop}%
\bibitem [{\citenamefont {Wang}\ and\ \citenamefont {Zaki}(2021)}]{wangStateEstimationTurbulent2021}%
  \BibitemOpen
  \bibfield  {author} {\bibinfo {author} {\bibfnamefont {M.}~\bibnamefont {Wang}}\ and\ \bibinfo {author} {\bibfnamefont {T.~A.}\ \bibnamefont {Zaki}},\ }\bibfield  {title} {\enquote {\bibinfo {title} {State estimation in turbulent channel flow from limited observations},}\ }\href {\doibase 10.1017/jfm.2021.268} {\bibfield  {journal} {\bibinfo  {journal} {Journal of Fluid Mechanics}\ }\textbf {\bibinfo {volume} {917}},\ \bibinfo {pages} {A9} (\bibinfo {year} {2021})}\BibitemShut {NoStop}%
\bibitem [{\citenamefont {Zaki}\ and\ \citenamefont {Wang}(2021)}]{zakiLimitedObservationsState2021}%
  \BibitemOpen
  \bibfield  {author} {\bibinfo {author} {\bibfnamefont {T.~A.}\ \bibnamefont {Zaki}}\ and\ \bibinfo {author} {\bibfnamefont {M.}~\bibnamefont {Wang}},\ }\bibfield  {title} {\enquote {\bibinfo {title} {From limited observations to the state of turbulence: {{Fundamental}} difficulties of flow reconstruction},}\ }\href {\doibase 10.1103/PhysRevFluids.6.100501} {\bibfield  {journal} {\bibinfo  {journal} {Physical Review Fluids}\ }\textbf {\bibinfo {volume} {6}},\ \bibinfo {pages} {100501} (\bibinfo {year} {2021})}\BibitemShut {NoStop}%
\bibitem [{\citenamefont {Di~Leoni}, \citenamefont {Mazzino},\ and\ \citenamefont {Biferale}(2020)}]{dileoniSynchronizationBigData2020}%
  \BibitemOpen
  \bibfield  {author} {\bibinfo {author} {\bibfnamefont {P.~C.}\ \bibnamefont {Di~Leoni}}, \bibinfo {author} {\bibfnamefont {A.}~\bibnamefont {Mazzino}}, \ and\ \bibinfo {author} {\bibfnamefont {L.}~\bibnamefont {Biferale}},\ }\bibfield  {title} {\enquote {\bibinfo {title} {Synchronization to {{Big Data}}: {{Nudging}} the {{Navier-Stokes Equations}} for {{Data Assimilation}} of {{Turbulent Flows}}},}\ }\href {\doibase 10.1103/PhysRevX.10.011023} {\bibfield  {journal} {\bibinfo  {journal} {Physical Review X}\ }\textbf {\bibinfo {volume} {10}},\ \bibinfo {pages} {011023} (\bibinfo {year} {2020})}\BibitemShut {NoStop}%
\bibitem [{\citenamefont {Zauner}\ \emph {et~al.}(2022)\citenamefont {Zauner}, \citenamefont {Mons}, \citenamefont {Marquet},\ and\ \citenamefont {Leclaire}}]{zaunerNudgingbasedDataAssimilation2022}%
  \BibitemOpen
  \bibfield  {author} {\bibinfo {author} {\bibfnamefont {M.}~\bibnamefont {Zauner}}, \bibinfo {author} {\bibfnamefont {V.}~\bibnamefont {Mons}}, \bibinfo {author} {\bibfnamefont {O.}~\bibnamefont {Marquet}}, \ and\ \bibinfo {author} {\bibfnamefont {B.}~\bibnamefont {Leclaire}},\ }\bibfield  {title} {\enquote {\bibinfo {title} {Nudging-based data assimilation of the turbulent flow around a square cylinder},}\ }\href {\doibase 10.1017/jfm.2022.133} {\bibfield  {journal} {\bibinfo  {journal} {Journal of Fluid Mechanics}\ }\textbf {\bibinfo {volume} {937}} (\bibinfo {year} {2022}),\ 10.1017/jfm.2022.133}\BibitemShut {NoStop}%
\bibitem [{\citenamefont {Fukami}, \citenamefont {Fukagata},\ and\ \citenamefont {Taira}(2023)}]{fukamiSuperresolutionAnalysisMachine2023}%
  \BibitemOpen
  \bibfield  {author} {\bibinfo {author} {\bibfnamefont {K.}~\bibnamefont {Fukami}}, \bibinfo {author} {\bibfnamefont {K.}~\bibnamefont {Fukagata}}, \ and\ \bibinfo {author} {\bibfnamefont {K.}~\bibnamefont {Taira}},\ }\bibfield  {title} {\enquote {\bibinfo {title} {Super-resolution analysis via machine learning: A survey for fluid flows},}\ }\href {\doibase 10.1007/s00162-023-00663-0} {\bibfield  {journal} {\bibinfo  {journal} {Theoretical and Computational Fluid Dynamics}\ } (\bibinfo {year} {2023}),\ 10.1007/s00162-023-00663-0}\BibitemShut {NoStop}%
\bibitem [{\citenamefont {Fukami}, \citenamefont {Fukagata},\ and\ \citenamefont {Taira}(2019)}]{fukamiSuperresolutionReconstructionTurbulent2019}%
  \BibitemOpen
  \bibfield  {author} {\bibinfo {author} {\bibfnamefont {K.}~\bibnamefont {Fukami}}, \bibinfo {author} {\bibfnamefont {K.}~\bibnamefont {Fukagata}}, \ and\ \bibinfo {author} {\bibfnamefont {K.}~\bibnamefont {Taira}},\ }\bibfield  {title} {\enquote {\bibinfo {title} {Super-resolution reconstruction of turbulent flows with machine learning},}\ }\href {\doibase 10.1017/jfm.2019.238} {\bibfield  {journal} {\bibinfo  {journal} {Journal of Fluid Mechanics}\ }\textbf {\bibinfo {volume} {870}},\ \bibinfo {pages} {106--120} (\bibinfo {year} {2019})}\BibitemShut {NoStop}%
\bibitem [{\citenamefont {Liu}\ \emph {et~al.}(2020)\citenamefont {Liu}, \citenamefont {Tang}, \citenamefont {Huang},\ and\ \citenamefont {Lu}}]{liuDeepLearningMethods2020}%
  \BibitemOpen
  \bibfield  {author} {\bibinfo {author} {\bibfnamefont {B.}~\bibnamefont {Liu}}, \bibinfo {author} {\bibfnamefont {J.}~\bibnamefont {Tang}}, \bibinfo {author} {\bibfnamefont {H.}~\bibnamefont {Huang}}, \ and\ \bibinfo {author} {\bibfnamefont {X.-Y.}\ \bibnamefont {Lu}},\ }\bibfield  {title} {\enquote {\bibinfo {title} {Deep learning methods for super-resolution reconstruction of turbulent flows},}\ }\href {\doibase 10.1063/1.5140772} {\bibfield  {journal} {\bibinfo  {journal} {Physics of Fluids}\ }\textbf {\bibinfo {volume} {32}},\ \bibinfo {pages} {025105} (\bibinfo {year} {2020})}\BibitemShut {NoStop}%
\bibitem [{\citenamefont {Fukami}, \citenamefont {Fukagata},\ and\ \citenamefont {Taira}(2021)}]{fukamiMachinelearningbasedSpatiotemporalSuper2021}%
  \BibitemOpen
  \bibfield  {author} {\bibinfo {author} {\bibfnamefont {K.}~\bibnamefont {Fukami}}, \bibinfo {author} {\bibfnamefont {K.}~\bibnamefont {Fukagata}}, \ and\ \bibinfo {author} {\bibfnamefont {K.}~\bibnamefont {Taira}},\ }\bibfield  {title} {\enquote {\bibinfo {title} {Machine-learning-based spatio-temporal super resolution reconstruction of turbulent flows},}\ }\href {\doibase 10.1017/jfm.2020.948} {\bibfield  {journal} {\bibinfo  {journal} {Journal of Fluid Mechanics}\ }\textbf {\bibinfo {volume} {909}} (\bibinfo {year} {2021}),\ 10.1017/jfm.2020.948}\BibitemShut {NoStop}%
\bibitem [{\citenamefont {Xie}\ \emph {et~al.}(2018)\citenamefont {Xie}, \citenamefont {Franz}, \citenamefont {Chu},\ and\ \citenamefont {Thuerey}}]{xieTempoGANTemporallyCoherent2018}%
  \BibitemOpen
  \bibfield  {author} {\bibinfo {author} {\bibfnamefont {Y.}~\bibnamefont {Xie}}, \bibinfo {author} {\bibfnamefont {E.}~\bibnamefont {Franz}}, \bibinfo {author} {\bibfnamefont {M.}~\bibnamefont {Chu}}, \ and\ \bibinfo {author} {\bibfnamefont {N.}~\bibnamefont {Thuerey}},\ }\bibfield  {title} {\enquote {\bibinfo {title} {{{tempoGAN}}: A temporally coherent, volumetric {{GAN}} for super-resolution fluid flow},}\ }\href {\doibase 10.1145/3197517.3201304} {\bibfield  {journal} {\bibinfo  {journal} {ACM Transactions on Graphics}\ }\textbf {\bibinfo {volume} {37}},\ \bibinfo {pages} {95:1--95:15} (\bibinfo {year} {2018})}\BibitemShut {NoStop}%
\bibitem [{\citenamefont {Deng}\ \emph {et~al.}(2019)\citenamefont {Deng}, \citenamefont {He}, \citenamefont {Liu},\ and\ \citenamefont {Kim}}]{dengSuperresolutionReconstructionTurbulent2019}%
  \BibitemOpen
  \bibfield  {author} {\bibinfo {author} {\bibfnamefont {Z.}~\bibnamefont {Deng}}, \bibinfo {author} {\bibfnamefont {C.}~\bibnamefont {He}}, \bibinfo {author} {\bibfnamefont {Y.}~\bibnamefont {Liu}}, \ and\ \bibinfo {author} {\bibfnamefont {K.~C.}\ \bibnamefont {Kim}},\ }\bibfield  {title} {\enquote {\bibinfo {title} {Super-resolution reconstruction of turbulent velocity fields using a generative adversarial network-based artificial intelligence framework},}\ }\href {\doibase 10.1063/1.5127031} {\bibfield  {journal} {\bibinfo  {journal} {Physics of Fluids}\ }\textbf {\bibinfo {volume} {31}},\ \bibinfo {pages} {125111} (\bibinfo {year} {2019})}\BibitemShut {NoStop}%
\bibitem [{\citenamefont {Stengel}\ \emph {et~al.}(2020)\citenamefont {Stengel}, \citenamefont {Glaws}, \citenamefont {Hettinger},\ and\ \citenamefont {King}}]{stengelAdversarialSuperresolutionClimatological2020}%
  \BibitemOpen
  \bibfield  {author} {\bibinfo {author} {\bibfnamefont {K.}~\bibnamefont {Stengel}}, \bibinfo {author} {\bibfnamefont {A.}~\bibnamefont {Glaws}}, \bibinfo {author} {\bibfnamefont {D.}~\bibnamefont {Hettinger}}, \ and\ \bibinfo {author} {\bibfnamefont {R.~N.}\ \bibnamefont {King}},\ }\bibfield  {title} {\enquote {\bibinfo {title} {Adversarial super-resolution of climatological wind and solar data},}\ }\href {\doibase 10.1073/pnas.1918964117} {\bibfield  {journal} {\bibinfo  {journal} {Proceedings of the National Academy of Sciences}\ }\textbf {\bibinfo {volume} {117}},\ \bibinfo {pages} {16805--16815} (\bibinfo {year} {2020})}\BibitemShut {NoStop}%
\bibitem [{\citenamefont {Yousif}, \citenamefont {Yu},\ and\ \citenamefont {Lim}(2021)}]{yousifHighfidelityReconstructionTurbulent2021}%
  \BibitemOpen
  \bibfield  {author} {\bibinfo {author} {\bibfnamefont {M.~Z.}\ \bibnamefont {Yousif}}, \bibinfo {author} {\bibfnamefont {L.}~\bibnamefont {Yu}}, \ and\ \bibinfo {author} {\bibfnamefont {H.-C.}\ \bibnamefont {Lim}},\ }\bibfield  {title} {\enquote {\bibinfo {title} {High-fidelity reconstruction of turbulent flow from spatially limited data using enhanced super-resolution generative adversarial network},}\ }\href {\doibase 10.1063/5.0066077} {\bibfield  {journal} {\bibinfo  {journal} {Physics of Fluids}\ }\textbf {\bibinfo {volume} {33}},\ \bibinfo {pages} {125119} (\bibinfo {year} {2021})}\BibitemShut {NoStop}%
\bibitem [{\citenamefont {Hassanaly}\ \emph {et~al.}(2022)\citenamefont {Hassanaly}, \citenamefont {Glaws}, \citenamefont {Stengel},\ and\ \citenamefont {King}}]{hassanalyAdversarialSamplingUnknown2022}%
  \BibitemOpen
  \bibfield  {author} {\bibinfo {author} {\bibfnamefont {M.}~\bibnamefont {Hassanaly}}, \bibinfo {author} {\bibfnamefont {A.}~\bibnamefont {Glaws}}, \bibinfo {author} {\bibfnamefont {K.}~\bibnamefont {Stengel}}, \ and\ \bibinfo {author} {\bibfnamefont {R.~N.}\ \bibnamefont {King}},\ }\bibfield  {title} {\enquote {\bibinfo {title} {Adversarial sampling of unknown and high-dimensional conditional distributions},}\ }\href {\doibase 10.1016/j.jcp.2021.110853} {\bibfield  {journal} {\bibinfo  {journal} {Journal of Computational Physics}\ }\textbf {\bibinfo {volume} {450}},\ \bibinfo {pages} {110853} (\bibinfo {year} {2022})}\BibitemShut {NoStop}%
\bibitem [{\citenamefont {Wang}\ \emph {et~al.}(2022)\citenamefont {Wang}, \citenamefont {Li}, \citenamefont {Liu}, \citenamefont {Wu}, \citenamefont {Hao}, \citenamefont {Zhang},\ and\ \citenamefont {He}}]{wangDeeplearningbasedSuperresolutionReconstruction2022}%
  \BibitemOpen
  \bibfield  {author} {\bibinfo {author} {\bibfnamefont {Z.}~\bibnamefont {Wang}}, \bibinfo {author} {\bibfnamefont {X.}~\bibnamefont {Li}}, \bibinfo {author} {\bibfnamefont {L.}~\bibnamefont {Liu}}, \bibinfo {author} {\bibfnamefont {X.}~\bibnamefont {Wu}}, \bibinfo {author} {\bibfnamefont {P.}~\bibnamefont {Hao}}, \bibinfo {author} {\bibfnamefont {X.}~\bibnamefont {Zhang}}, \ and\ \bibinfo {author} {\bibfnamefont {F.}~\bibnamefont {He}},\ }\bibfield  {title} {\enquote {\bibinfo {title} {Deep-learning-based super-resolution reconstruction of high-speed imaging in fluids},}\ }\href {\doibase 10.1063/5.0078644} {\bibfield  {journal} {\bibinfo  {journal} {Physics of Fluids}\ }\textbf {\bibinfo {volume} {34}},\ \bibinfo {pages} {037107} (\bibinfo {year} {2022})}\BibitemShut {NoStop}%
\bibitem [{\citenamefont {Buzzicotti}\ \emph {et~al.}(2021)\citenamefont {Buzzicotti}, \citenamefont {Bonaccorso}, \citenamefont {Di~Leoni},\ and\ \citenamefont {Biferale}}]{buzzicottiReconstructionTurbulentData2021}%
  \BibitemOpen
  \bibfield  {author} {\bibinfo {author} {\bibfnamefont {M.}~\bibnamefont {Buzzicotti}}, \bibinfo {author} {\bibfnamefont {F.}~\bibnamefont {Bonaccorso}}, \bibinfo {author} {\bibfnamefont {P.~C.}\ \bibnamefont {Di~Leoni}}, \ and\ \bibinfo {author} {\bibfnamefont {L.}~\bibnamefont {Biferale}},\ }\bibfield  {title} {\enquote {\bibinfo {title} {Reconstruction of turbulent data with deep generative models for semantic inpainting from {{TURB-Rot}} database},}\ }\href {\doibase 10.1103/PhysRevFluids.6.050503} {\bibfield  {journal} {\bibinfo  {journal} {Physical Review Fluids}\ }\textbf {\bibinfo {volume} {6}},\ \bibinfo {pages} {050503} (\bibinfo {year} {2021})}\BibitemShut {NoStop}%
\bibitem [{\citenamefont {Gundersen}\ \emph {et~al.}(2021)\citenamefont {Gundersen}, \citenamefont {Oleynik}, \citenamefont {Blaser},\ and\ \citenamefont {Alendal}}]{gundersenSemiconditionalVariationalAutoencoder2021}%
  \BibitemOpen
  \bibfield  {author} {\bibinfo {author} {\bibfnamefont {K.}~\bibnamefont {Gundersen}}, \bibinfo {author} {\bibfnamefont {A.}~\bibnamefont {Oleynik}}, \bibinfo {author} {\bibfnamefont {N.}~\bibnamefont {Blaser}}, \ and\ \bibinfo {author} {\bibfnamefont {G.}~\bibnamefont {Alendal}},\ }\bibfield  {title} {\enquote {\bibinfo {title} {Semi-conditional variational auto-encoder for flow reconstruction and uncertainty quantification from limited observations},}\ }\href {\doibase 10.1063/5.0025779} {\bibfield  {journal} {\bibinfo  {journal} {Physics of Fluids}\ }\textbf {\bibinfo {volume} {33}},\ \bibinfo {pages} {017119} (\bibinfo {year} {2021})}\BibitemShut {NoStop}%
\bibitem [{\citenamefont {Cai}\ \emph {et~al.}(2021)\citenamefont {Cai}, \citenamefont {Wang}, \citenamefont {Fuest}, \citenamefont {Jeon}, \citenamefont {Gray},\ and\ \citenamefont {Karniadakis}}]{caiFlowEspressoCup2021}%
  \BibitemOpen
  \bibfield  {author} {\bibinfo {author} {\bibfnamefont {S.}~\bibnamefont {Cai}}, \bibinfo {author} {\bibfnamefont {Z.}~\bibnamefont {Wang}}, \bibinfo {author} {\bibfnamefont {F.}~\bibnamefont {Fuest}}, \bibinfo {author} {\bibfnamefont {Y.~J.}\ \bibnamefont {Jeon}}, \bibinfo {author} {\bibfnamefont {C.}~\bibnamefont {Gray}}, \ and\ \bibinfo {author} {\bibfnamefont {G.~E.}\ \bibnamefont {Karniadakis}},\ }\bibfield  {title} {\enquote {\bibinfo {title} {Flow over an espresso cup: Inferring 3-{{D}} velocity and pressure fields from tomographic background oriented {{Schlieren}} via physics-informed neural networks},}\ }\href {\doibase 10.1017/jfm.2021.135} {\bibfield  {journal} {\bibinfo  {journal} {Journal of Fluid Mechanics}\ }\textbf {\bibinfo {volume} {915}},\ \bibinfo {pages} {A102} (\bibinfo {year} {2021})}\BibitemShut {NoStop}%
\bibitem [{\citenamefont {Clark Di~Leoni}\ \emph {et~al.}(2023{\natexlab{a}})\citenamefont {Clark Di~Leoni}, \citenamefont {Agarwal}, \citenamefont {Zaki}, \citenamefont {Meneveau},\ and\ \citenamefont {Katz}}]{clarkdileoniReconstructingTurbulentVelocity2023}%
  \BibitemOpen
  \bibfield  {author} {\bibinfo {author} {\bibfnamefont {P.}~\bibnamefont {Clark Di~Leoni}}, \bibinfo {author} {\bibfnamefont {K.}~\bibnamefont {Agarwal}}, \bibinfo {author} {\bibfnamefont {T.~A.}\ \bibnamefont {Zaki}}, \bibinfo {author} {\bibfnamefont {C.}~\bibnamefont {Meneveau}}, \ and\ \bibinfo {author} {\bibfnamefont {J.}~\bibnamefont {Katz}},\ }\bibfield  {title} {\enquote {\bibinfo {title} {Reconstructing turbulent velocity and pressure fields from under-resolved noisy particle tracks using physics-informed neural networks},}\ }\href {\doibase 10.1007/s00348-023-03629-4} {\bibfield  {journal} {\bibinfo  {journal} {Experiments in Fluids}\ }\textbf {\bibinfo {volume} {64}},\ \bibinfo {pages} {95} (\bibinfo {year} {2023}{\natexlab{a}})}\BibitemShut {NoStop}%
\bibitem [{\citenamefont {Du}, \citenamefont {Wang},\ and\ \citenamefont {Zaki}(2022)}]{duStateEstimationMinimal2022}%
  \BibitemOpen
  \bibfield  {author} {\bibinfo {author} {\bibfnamefont {Y.}~\bibnamefont {Du}}, \bibinfo {author} {\bibfnamefont {M.}~\bibnamefont {Wang}}, \ and\ \bibinfo {author} {\bibfnamefont {T.~A.}\ \bibnamefont {Zaki}},\ }\href@noop {} {\enquote {\bibinfo {title} {State estimation in minimal turbulent channel flow: {{A}} comparative study of {{4DVar}} and {{PINN}}},}\ } (\bibinfo {year} {2022}),\ \Eprint {http://arxiv.org/abs/2210.09424} {arxiv:2210.09424 [physics]} \BibitemShut {NoStop}%
\bibitem [{\citenamefont {Clark Di~Leoni}\ \emph {et~al.}(2023{\natexlab{b}})\citenamefont {Clark Di~Leoni}, \citenamefont {Lu}, \citenamefont {Meneveau}, \citenamefont {Karniadakis},\ and\ \citenamefont {Zaki}}]{clarkdileoniNeuralOperatorPrediction2023}%
  \BibitemOpen
  \bibfield  {author} {\bibinfo {author} {\bibfnamefont {P.}~\bibnamefont {Clark Di~Leoni}}, \bibinfo {author} {\bibfnamefont {L.}~\bibnamefont {Lu}}, \bibinfo {author} {\bibfnamefont {C.}~\bibnamefont {Meneveau}}, \bibinfo {author} {\bibfnamefont {G.~E.}\ \bibnamefont {Karniadakis}}, \ and\ \bibinfo {author} {\bibfnamefont {T.~A.}\ \bibnamefont {Zaki}},\ }\bibfield  {title} {\enquote {\bibinfo {title} {Neural operator prediction of linear instability waves in high-speed boundary layers},}\ }\href {\doibase 10.1016/j.jcp.2022.111793} {\bibfield  {journal} {\bibinfo  {journal} {Journal of Computational Physics}\ }\textbf {\bibinfo {volume} {474}},\ \bibinfo {pages} {111793} (\bibinfo {year} {2023}{\natexlab{b}})}\BibitemShut {NoStop}%
\bibitem [{\citenamefont {Yousif}\ \emph {et~al.}(2023)\citenamefont {Yousif}, \citenamefont {Yu}, \citenamefont {Hoyas}, \citenamefont {Vinuesa},\ and\ \citenamefont {Lim}}]{yousifDeeplearningApproachReconstructing2023}%
  \BibitemOpen
  \bibfield  {author} {\bibinfo {author} {\bibfnamefont {M.~Z.}\ \bibnamefont {Yousif}}, \bibinfo {author} {\bibfnamefont {L.}~\bibnamefont {Yu}}, \bibinfo {author} {\bibfnamefont {S.}~\bibnamefont {Hoyas}}, \bibinfo {author} {\bibfnamefont {R.}~\bibnamefont {Vinuesa}}, \ and\ \bibinfo {author} {\bibfnamefont {H.}~\bibnamefont {Lim}},\ }\bibfield  {title} {\enquote {\bibinfo {title} {A deep-learning approach for reconstructing {{3D}} turbulent flows from {{2D}} observation data},}\ }\href {\doibase 10.1038/s41598-023-29525-9} {\bibfield  {journal} {\bibinfo  {journal} {Scientific Reports}\ }\textbf {\bibinfo {volume} {13}},\ \bibinfo {pages} {2529} (\bibinfo {year} {2023})}\BibitemShut {NoStop}%
\bibitem [{\citenamefont {Li}\ \emph {et~al.}(2023)\citenamefont {Li}, \citenamefont {Zheng}, \citenamefont {Kovachki}, \citenamefont {Jin}, \citenamefont {Chen}, \citenamefont {Liu}, \citenamefont {Azizzadenesheli},\ and\ \citenamefont {Anandkumar}}]{liPhysicsInformedNeuralOperator2023}%
  \BibitemOpen
  \bibfield  {author} {\bibinfo {author} {\bibfnamefont {Z.}~\bibnamefont {Li}}, \bibinfo {author} {\bibfnamefont {H.}~\bibnamefont {Zheng}}, \bibinfo {author} {\bibfnamefont {N.}~\bibnamefont {Kovachki}}, \bibinfo {author} {\bibfnamefont {D.}~\bibnamefont {Jin}}, \bibinfo {author} {\bibfnamefont {H.}~\bibnamefont {Chen}}, \bibinfo {author} {\bibfnamefont {B.}~\bibnamefont {Liu}}, \bibinfo {author} {\bibfnamefont {K.}~\bibnamefont {Azizzadenesheli}}, \ and\ \bibinfo {author} {\bibfnamefont {A.}~\bibnamefont {Anandkumar}},\ }\href@noop {} {\enquote {\bibinfo {title} {Physics-{{Informed Neural Operator}} for {{Learning Partial Differential Equations}}},}\ } (\bibinfo {year} {2023}),\ \Eprint {http://arxiv.org/abs/2111.03794} {arxiv:2111.03794 [cs, math]} \BibitemShut {NoStop}%
\bibitem [{\citenamefont {Wilks}(2019)}]{wilksStatisticalMethodsAtmospheric2019}%
  \BibitemOpen
  \bibfield  {author} {\bibinfo {author} {\bibfnamefont {D.~S.}\ \bibnamefont {Wilks}},\ }\bibfield  {title} {\enquote {\bibinfo {title} {Statistical {{Methods}} in the {{Atmospheric Sciences}}},}\ }in\ \href {\doibase 10.1016/B978-0-12-815823-4.09987-9} {\emph {\bibinfo {booktitle} {Statistical {{Methods}} in the {{Atmospheric Sciences}} ({{Fourth Edition}})}}}\ (\bibinfo  {publisher} {{Elsevier}},\ \bibinfo {year} {2019})\ pp.\ \bibinfo {pages} {i--ii}\BibitemShut {NoStop}%
\bibitem [{\citenamefont {Salimans}\ \emph {et~al.}(2016)\citenamefont {Salimans}, \citenamefont {Goodfellow}, \citenamefont {Zaremba}, \citenamefont {Cheung}, \citenamefont {Radford}, \citenamefont {Chen},\ and\ \citenamefont {Chen}}]{salimansImprovedTechniquesTraining2016}%
  \BibitemOpen
  \bibfield  {author} {\bibinfo {author} {\bibfnamefont {T.}~\bibnamefont {Salimans}}, \bibinfo {author} {\bibfnamefont {I.}~\bibnamefont {Goodfellow}}, \bibinfo {author} {\bibfnamefont {W.}~\bibnamefont {Zaremba}}, \bibinfo {author} {\bibfnamefont {V.}~\bibnamefont {Cheung}}, \bibinfo {author} {\bibfnamefont {A.}~\bibnamefont {Radford}}, \bibinfo {author} {\bibfnamefont {X.}~\bibnamefont {Chen}}, \ and\ \bibinfo {author} {\bibfnamefont {X.}~\bibnamefont {Chen}},\ }\bibfield  {title} {\enquote {\bibinfo {title} {Improved {{Techniques}} for {{Training GANs}}},}\ }in\ \href@noop {} {\emph {\bibinfo {booktitle} {Advances in {{Neural Information Processing Systems}}}}},\ Vol.~\bibinfo {volume} {29}\ (\bibinfo  {publisher} {{Curran Associates, Inc.}},\ \bibinfo {year} {2016})\BibitemShut {NoStop}%
\bibitem [{\citenamefont {{Sohl-Dickstein}}\ \emph {et~al.}(2015)\citenamefont {{Sohl-Dickstein}}, \citenamefont {Weiss}, \citenamefont {Maheswaranathan},\ and\ \citenamefont {Ganguli}}]{sohl-dicksteinDeepUnsupervisedLearning2015}%
  \BibitemOpen
  \bibfield  {author} {\bibinfo {author} {\bibfnamefont {J.}~\bibnamefont {{Sohl-Dickstein}}}, \bibinfo {author} {\bibfnamefont {E.}~\bibnamefont {Weiss}}, \bibinfo {author} {\bibfnamefont {N.}~\bibnamefont {Maheswaranathan}}, \ and\ \bibinfo {author} {\bibfnamefont {S.}~\bibnamefont {Ganguli}},\ }\bibfield  {title} {\enquote {\bibinfo {title} {Deep {{Unsupervised Learning}} using {{Nonequilibrium Thermodynamics}}},}\ }in\ \href@noop {} {\emph {\bibinfo {booktitle} {Proceedings of the 32nd {{International Conference}} on {{Machine Learning}}}}}\ (\bibinfo  {publisher} {{PMLR}},\ \bibinfo {year} {2015})\ pp.\ \bibinfo {pages} {2256--2265}\BibitemShut {NoStop}%
\bibitem [{\citenamefont {Song}\ and\ \citenamefont {Ermon}(2020)}]{songGenerativeModelingEstimating2020}%
  \BibitemOpen
  \bibfield  {author} {\bibinfo {author} {\bibfnamefont {Y.}~\bibnamefont {Song}}\ and\ \bibinfo {author} {\bibfnamefont {S.}~\bibnamefont {Ermon}},\ }\href {\doibase 10.48550/arXiv.1907.05600} {\enquote {\bibinfo {title} {Generative {{Modeling}} by {{Estimating Gradients}} of the {{Data Distribution}}},}\ } (\bibinfo {year} {2020}),\ \Eprint {http://arxiv.org/abs/1907.05600} {arxiv:1907.05600 [cs, stat]} \BibitemShut {NoStop}%
\bibitem [{\citenamefont {Dhariwal}\ and\ \citenamefont {Nichol}(2021)}]{dhariwalDiffusionModelsBeat2021}%
  \BibitemOpen
  \bibfield  {author} {\bibinfo {author} {\bibfnamefont {P.}~\bibnamefont {Dhariwal}}\ and\ \bibinfo {author} {\bibfnamefont {A.}~\bibnamefont {Nichol}},\ }\bibfield  {title} {\enquote {\bibinfo {title} {Diffusion {{Models Beat GANs}} on {{Image Synthesis}}},}\ }in\ \href@noop {} {\emph {\bibinfo {booktitle} {Advances in {{Neural Information Processing Systems}}}}},\ Vol.~\bibinfo {volume} {34}\ (\bibinfo  {publisher} {{Curran Associates, Inc.}},\ \bibinfo {year} {2021})\ pp.\ \bibinfo {pages} {8780--8794}\BibitemShut {NoStop}%
\bibitem [{\citenamefont {Rombach}\ \emph {et~al.}(2022)\citenamefont {Rombach}, \citenamefont {Blattmann}, \citenamefont {Lorenz}, \citenamefont {Esser},\ and\ \citenamefont {Ommer}}]{rombachHighResolutionImageSynthesis2022}%
  \BibitemOpen
  \bibfield  {author} {\bibinfo {author} {\bibfnamefont {R.}~\bibnamefont {Rombach}}, \bibinfo {author} {\bibfnamefont {A.}~\bibnamefont {Blattmann}}, \bibinfo {author} {\bibfnamefont {D.}~\bibnamefont {Lorenz}}, \bibinfo {author} {\bibfnamefont {P.}~\bibnamefont {Esser}}, \ and\ \bibinfo {author} {\bibfnamefont {B.}~\bibnamefont {Ommer}},\ }\href {\doibase 10.48550/arXiv.2112.10752} {\enquote {\bibinfo {title} {High-{{Resolution Image Synthesis}} with {{Latent Diffusion Models}}},}\ } (\bibinfo {year} {2022}),\ \Eprint {http://arxiv.org/abs/2112.10752} {arxiv:2112.10752 [cs]} \BibitemShut {NoStop}%
\bibitem [{\citenamefont {Pinaya}\ \emph {et~al.}(2022)\citenamefont {Pinaya}, \citenamefont {Tudosiu}, \citenamefont {Dafflon}, \citenamefont {{da Costa}}, \citenamefont {Fernandez}, \citenamefont {Nachev}, \citenamefont {Ourselin},\ and\ \citenamefont {Cardoso}}]{pinayaBrainImagingGeneration2022}%
  \BibitemOpen
  \bibfield  {author} {\bibinfo {author} {\bibfnamefont {W.~H.~L.}\ \bibnamefont {Pinaya}}, \bibinfo {author} {\bibfnamefont {P.-D.}\ \bibnamefont {Tudosiu}}, \bibinfo {author} {\bibfnamefont {J.}~\bibnamefont {Dafflon}}, \bibinfo {author} {\bibfnamefont {P.~F.}\ \bibnamefont {{da Costa}}}, \bibinfo {author} {\bibfnamefont {V.}~\bibnamefont {Fernandez}}, \bibinfo {author} {\bibfnamefont {P.}~\bibnamefont {Nachev}}, \bibinfo {author} {\bibfnamefont {S.}~\bibnamefont {Ourselin}}, \ and\ \bibinfo {author} {\bibfnamefont {M.~J.}\ \bibnamefont {Cardoso}},\ }\href {\doibase 10.48550/arXiv.2209.07162} {\enquote {\bibinfo {title} {Brain {{Imaging Generation}} with {{Latent Diffusion Models}}},}\ } (\bibinfo {year} {2022}),\ \Eprint {http://arxiv.org/abs/2209.07162} {arxiv:2209.07162 [cs, eess, q-bio]} \BibitemShut {NoStop}%
\bibitem [{\citenamefont {Lugmayr}\ \emph {et~al.}(2022)\citenamefont {Lugmayr}, \citenamefont {Danelljan}, \citenamefont {Romero}, \citenamefont {Yu}, \citenamefont {Timofte},\ and\ \citenamefont {Van~Gool}}]{lugmayrRePaintInpaintingUsing2022}%
  \BibitemOpen
  \bibfield  {author} {\bibinfo {author} {\bibfnamefont {A.}~\bibnamefont {Lugmayr}}, \bibinfo {author} {\bibfnamefont {M.}~\bibnamefont {Danelljan}}, \bibinfo {author} {\bibfnamefont {A.}~\bibnamefont {Romero}}, \bibinfo {author} {\bibfnamefont {F.}~\bibnamefont {Yu}}, \bibinfo {author} {\bibfnamefont {R.}~\bibnamefont {Timofte}}, \ and\ \bibinfo {author} {\bibfnamefont {L.}~\bibnamefont {Van~Gool}},\ }\bibfield  {title} {\enquote {\bibinfo {title} {{{RePaint}}: {{Inpainting Using Denoising Diffusion Probabilistic Models}}},}\ }in\ \href@noop {} {\emph {\bibinfo {booktitle} {Proceedings of the {{IEEE}}/{{CVF Conference}} on {{Computer Vision}} and {{Pattern Recognition}}}}}\ (\bibinfo {year} {2022})\ pp.\ \bibinfo {pages} {11461--11471}\BibitemShut {NoStop}%
\bibitem [{\citenamefont {Saharia}\ \emph {et~al.}(2022)\citenamefont {Saharia}, \citenamefont {Chan}, \citenamefont {Chang}, \citenamefont {Lee}, \citenamefont {Ho}, \citenamefont {Salimans}, \citenamefont {Fleet},\ and\ \citenamefont {Norouzi}}]{sahariaPaletteImagetoImageDiffusion2022}%
  \BibitemOpen
  \bibfield  {author} {\bibinfo {author} {\bibfnamefont {C.}~\bibnamefont {Saharia}}, \bibinfo {author} {\bibfnamefont {W.}~\bibnamefont {Chan}}, \bibinfo {author} {\bibfnamefont {H.}~\bibnamefont {Chang}}, \bibinfo {author} {\bibfnamefont {C.~A.}\ \bibnamefont {Lee}}, \bibinfo {author} {\bibfnamefont {J.}~\bibnamefont {Ho}}, \bibinfo {author} {\bibfnamefont {T.}~\bibnamefont {Salimans}}, \bibinfo {author} {\bibfnamefont {D.~J.}\ \bibnamefont {Fleet}}, \ and\ \bibinfo {author} {\bibfnamefont {M.}~\bibnamefont {Norouzi}},\ }\href@noop {} {\enquote {\bibinfo {title} {Palette: {{Image-to-Image Diffusion Models}}},}\ } (\bibinfo {year} {2022}),\ \Eprint {http://arxiv.org/abs/2111.05826} {arxiv:2111.05826 [cs]} \BibitemShut {NoStop}%
\bibitem [{\citenamefont {Courtier}, \citenamefont {Th{\'e}paut},\ and\ \citenamefont {Hollingsworth}(1994)}]{courtierStrategyOperationalImplementation1994}%
  \BibitemOpen
  \bibfield  {author} {\bibinfo {author} {\bibfnamefont {P.}~\bibnamefont {Courtier}}, \bibinfo {author} {\bibfnamefont {J.-N.}\ \bibnamefont {Th{\'e}paut}}, \ and\ \bibinfo {author} {\bibfnamefont {A.}~\bibnamefont {Hollingsworth}},\ }\bibfield  {title} {\enquote {\bibinfo {title} {A strategy for operational implementation of {{4D-Var}}, using an incremental approach},}\ }\href {\doibase 10.1002/qj.49712051912} {\bibfield  {journal} {\bibinfo  {journal} {Quarterly Journal of the Royal Meteorological Society}\ }\textbf {\bibinfo {volume} {120}},\ \bibinfo {pages} {1367--1387} (\bibinfo {year} {1994})}\BibitemShut {NoStop}%
\bibitem [{\citenamefont {Evensen}(1994)}]{evensenSequentialDataAssimilation1994}%
  \BibitemOpen
  \bibfield  {author} {\bibinfo {author} {\bibfnamefont {G.}~\bibnamefont {Evensen}},\ }\bibfield  {title} {\enquote {\bibinfo {title} {Sequential data assimilation with a nonlinear quasi-geostrophic model using {{Monte Carlo}} methods to forecast error statistics},}\ }\href {\doibase 10.1029/94JC00572} {\bibfield  {journal} {\bibinfo  {journal} {Journal of Geophysical Research: Oceans}\ }\textbf {\bibinfo {volume} {99}},\ \bibinfo {pages} {10143--10162} (\bibinfo {year} {1994})}\BibitemShut {NoStop}%
\bibitem [{\citenamefont {Sprague}(2021)}]{sprague2021exawind}%
  \BibitemOpen
  \bibfield  {author} {\bibinfo {author} {\bibfnamefont {M.~A.}\ \bibnamefont {Sprague}},\ }\href@noop {} {\enquote {\bibinfo {title} {{{ExaWind}}: {{Predictive}} wind energy simulations},}\ }\bibinfo {type} {Tech. Rep.}\ (\bibinfo  {institution} {{National Renewable Energy Lab.(NREL), Golden, CO (United States)}},\ \bibinfo {year} {2021})\BibitemShut {NoStop}%
\bibitem [{\citenamefont {Almgren}\ \emph {et~al.}(1998)\citenamefont {Almgren}, \citenamefont {Bell}, \citenamefont {Colella}, \citenamefont {Howell},\ and\ \citenamefont {Welcome}}]{almgrenConservativeAdaptiveProjection1998}%
  \BibitemOpen
  \bibfield  {author} {\bibinfo {author} {\bibfnamefont {A.~S.}\ \bibnamefont {Almgren}}, \bibinfo {author} {\bibfnamefont {J.~B.}\ \bibnamefont {Bell}}, \bibinfo {author} {\bibfnamefont {P.}~\bibnamefont {Colella}}, \bibinfo {author} {\bibfnamefont {L.~H.}\ \bibnamefont {Howell}}, \ and\ \bibinfo {author} {\bibfnamefont {M.~L.}\ \bibnamefont {Welcome}},\ }\bibfield  {title} {\enquote {\bibinfo {title} {A {{Conservative Adaptive Projection Method}} for the {{Variable Density Incompressible Navier}}{\textendash}{{Stokes Equations}}},}\ }\href {\doibase 10.1006/jcph.1998.5890} {\bibfield  {journal} {\bibinfo  {journal} {Journal of Computational Physics}\ }\textbf {\bibinfo {volume} {142}},\ \bibinfo {pages} {1--46} (\bibinfo {year} {1998})}\BibitemShut {NoStop}%
\bibitem [{\citenamefont {Smagorinsky}(1963)}]{smagorinskyGENERALCIRCULATIONEXPERIMENTS1963}%
  \BibitemOpen
  \bibfield  {author} {\bibinfo {author} {\bibfnamefont {J.}~\bibnamefont {Smagorinsky}},\ }\bibfield  {title} {\enquote {\bibinfo {title} {{{GENERAL CIRCULATION EXPERIMENTS WITH THE PRIMITIVE EQUATIONS}}: {{I}}. {{THE BASIC EXPERIMENT}}},}\ }\href {\doibase 10.1175/1520-0493(1963)091<0099:GCEWTP>2.3.CO;2} {\bibfield  {journal} {\bibinfo  {journal} {Monthly Weather Review}\ }\textbf {\bibinfo {volume} {91}},\ \bibinfo {pages} {99--164} (\bibinfo {year} {1963})}\BibitemShut {NoStop}%
\bibitem [{\citenamefont {Luo}(2022)}]{luoUnderstandingDiffusionModels2022}%
  \BibitemOpen
  \bibfield  {author} {\bibinfo {author} {\bibfnamefont {C.}~\bibnamefont {Luo}},\ }\href {\doibase 10.48550/arXiv.2208.11970} {\enquote {\bibinfo {title} {Understanding {{Diffusion Models}}: {{A Unified Perspective}}},}\ } (\bibinfo {year} {2022}),\ \Eprint {http://arxiv.org/abs/2208.11970} {arxiv:2208.11970 [cs]} \BibitemShut {NoStop}%
\bibitem [{\citenamefont {Song}\ \emph {et~al.}(2021)\citenamefont {Song}, \citenamefont {{Sohl-Dickstein}}, \citenamefont {Kingma}, \citenamefont {Kumar}, \citenamefont {Ermon},\ and\ \citenamefont {Poole}}]{songScoreBasedGenerativeModeling2021}%
  \BibitemOpen
  \bibfield  {author} {\bibinfo {author} {\bibfnamefont {Y.}~\bibnamefont {Song}}, \bibinfo {author} {\bibfnamefont {J.}~\bibnamefont {{Sohl-Dickstein}}}, \bibinfo {author} {\bibfnamefont {D.~P.}\ \bibnamefont {Kingma}}, \bibinfo {author} {\bibfnamefont {A.}~\bibnamefont {Kumar}}, \bibinfo {author} {\bibfnamefont {S.}~\bibnamefont {Ermon}}, \ and\ \bibinfo {author} {\bibfnamefont {B.}~\bibnamefont {Poole}},\ }\href {\doibase 10.48550/arXiv.2011.13456} {\enquote {\bibinfo {title} {Score-{{Based Generative Modeling}} through {{Stochastic Differential Equations}}},}\ } (\bibinfo {year} {2021}),\ \Eprint {http://arxiv.org/abs/2011.13456} {arxiv:2011.13456 [cs, stat]} \BibitemShut {NoStop}%
\bibitem [{\citenamefont {Kawar}\ \emph {et~al.}(2022)\citenamefont {Kawar}, \citenamefont {Elad}, \citenamefont {Ermon},\ and\ \citenamefont {Song}}]{kawarDenoisingDiffusionRestoration2022}%
  \BibitemOpen
  \bibfield  {author} {\bibinfo {author} {\bibfnamefont {B.}~\bibnamefont {Kawar}}, \bibinfo {author} {\bibfnamefont {M.}~\bibnamefont {Elad}}, \bibinfo {author} {\bibfnamefont {S.}~\bibnamefont {Ermon}}, \ and\ \bibinfo {author} {\bibfnamefont {J.}~\bibnamefont {Song}},\ }\bibfield  {title} {\enquote {\bibinfo {title} {Denoising {{Diffusion Restoration Models}}},}\ }\href {\doibase 10.48550/arXiv.2201.11793} {\  (\bibinfo {year} {2022}),\ 10.48550/arXiv.2201.11793}\BibitemShut {NoStop}%
\bibitem [{\citenamefont {Kingma}\ and\ \citenamefont {Welling}(2013)}]{kingmaAutoEncodingVariationalBayes2013}%
  \BibitemOpen
  \bibfield  {author} {\bibinfo {author} {\bibfnamefont {D.~P.}\ \bibnamefont {Kingma}}\ and\ \bibinfo {author} {\bibfnamefont {M.}~\bibnamefont {Welling}},\ }\bibfield  {title} {\enquote {\bibinfo {title} {Auto-{{Encoding Variational Bayes}}},}\ }\href {\doibase 10.48550/arXiv.1312.6114} {\  (\bibinfo {year} {2013}),\ 10.48550/arXiv.1312.6114}\BibitemShut {NoStop}%
\bibitem [{\citenamefont {Zhang}\ \emph {et~al.}(2018)\citenamefont {Zhang}, \citenamefont {Isola}, \citenamefont {Efros}, \citenamefont {Shechtman},\ and\ \citenamefont {Wang}}]{zhangUnreasonableEffectivenessDeep2018}%
  \BibitemOpen
  \bibfield  {author} {\bibinfo {author} {\bibfnamefont {R.}~\bibnamefont {Zhang}}, \bibinfo {author} {\bibfnamefont {P.}~\bibnamefont {Isola}}, \bibinfo {author} {\bibfnamefont {A.~A.}\ \bibnamefont {Efros}}, \bibinfo {author} {\bibfnamefont {E.}~\bibnamefont {Shechtman}}, \ and\ \bibinfo {author} {\bibfnamefont {O.}~\bibnamefont {Wang}},\ }\href {\doibase 10.48550/arXiv.1801.03924} {\enquote {\bibinfo {title} {The {{Unreasonable Effectiveness}} of {{Deep Features}} as a {{Perceptual Metric}}},}\ } (\bibinfo {year} {2018}),\ \Eprint {http://arxiv.org/abs/1801.03924} {arxiv:1801.03924 [cs]} \BibitemShut {NoStop}%
\bibitem [{\citenamefont {Isola}\ \emph {et~al.}(2018)\citenamefont {Isola}, \citenamefont {Zhu}, \citenamefont {Zhou},\ and\ \citenamefont {Efros}}]{isolaImagetoImageTranslationConditional2018}%
  \BibitemOpen
  \bibfield  {author} {\bibinfo {author} {\bibfnamefont {P.}~\bibnamefont {Isola}}, \bibinfo {author} {\bibfnamefont {J.-Y.}\ \bibnamefont {Zhu}}, \bibinfo {author} {\bibfnamefont {T.}~\bibnamefont {Zhou}}, \ and\ \bibinfo {author} {\bibfnamefont {A.~A.}\ \bibnamefont {Efros}},\ }\href {\doibase 10.48550/arXiv.1611.07004} {\enquote {\bibinfo {title} {Image-to-{{Image Translation}} with {{Conditional Adversarial Networks}}},}\ } (\bibinfo {year} {2018}),\ \Eprint {http://arxiv.org/abs/1611.07004} {arxiv:1611.07004 [cs]} \BibitemShut {NoStop}%
\bibitem [{\citenamefont {Brock}, \citenamefont {Donahue},\ and\ \citenamefont {Simonyan}(2019)}]{brockLargeScaleGAN2019}%
  \BibitemOpen
  \bibfield  {author} {\bibinfo {author} {\bibfnamefont {A.}~\bibnamefont {Brock}}, \bibinfo {author} {\bibfnamefont {J.}~\bibnamefont {Donahue}}, \ and\ \bibinfo {author} {\bibfnamefont {K.}~\bibnamefont {Simonyan}},\ }\href {\doibase 10.48550/arXiv.1809.11096} {\enquote {\bibinfo {title} {Large {{Scale GAN Training}} for {{High Fidelity Natural Image Synthesis}}},}\ } (\bibinfo {year} {2019}),\ \Eprint {http://arxiv.org/abs/1809.11096} {arxiv:1809.11096 [cs, stat]} \BibitemShut {NoStop}%
\bibitem [{\citenamefont {Wu}\ and\ \citenamefont {He}(2018)}]{wuGroupNormalization2018}%
  \BibitemOpen
  \bibfield  {author} {\bibinfo {author} {\bibfnamefont {Y.}~\bibnamefont {Wu}}\ and\ \bibinfo {author} {\bibfnamefont {K.}~\bibnamefont {He}},\ }\href {\doibase 10.48550/arXiv.1803.08494} {\enquote {\bibinfo {title} {Group {{Normalization}}},}\ } (\bibinfo {year} {2018}),\ \Eprint {http://arxiv.org/abs/1803.08494} {arxiv:1803.08494 [cs]} \BibitemShut {NoStop}%
\bibitem [{\citenamefont {Ulyanov}, \citenamefont {Vedaldi},\ and\ \citenamefont {Lempitsky}(2017)}]{ulyanovInstanceNormalizationMissing2017}%
  \BibitemOpen
  \bibfield  {author} {\bibinfo {author} {\bibfnamefont {D.}~\bibnamefont {Ulyanov}}, \bibinfo {author} {\bibfnamefont {A.}~\bibnamefont {Vedaldi}}, \ and\ \bibinfo {author} {\bibfnamefont {V.}~\bibnamefont {Lempitsky}},\ }\href {\doibase 10.48550/arXiv.1607.08022} {\enquote {\bibinfo {title} {Instance {{Normalization}}: {{The Missing Ingredient}} for {{Fast Stylization}}},}\ } (\bibinfo {year} {2017}),\ \Eprint {http://arxiv.org/abs/1607.08022} {arxiv:1607.08022 [cs]} \BibitemShut {NoStop}%
\bibitem [{\citenamefont {Carrassi}\ \emph {et~al.}(2018)\citenamefont {Carrassi}, \citenamefont {Bocquet}, \citenamefont {Bertino},\ and\ \citenamefont {Evensen}}]{carrassiDataAssimilationGeosciences2018}%
  \BibitemOpen
  \bibfield  {author} {\bibinfo {author} {\bibfnamefont {A.}~\bibnamefont {Carrassi}}, \bibinfo {author} {\bibfnamefont {M.}~\bibnamefont {Bocquet}}, \bibinfo {author} {\bibfnamefont {L.}~\bibnamefont {Bertino}}, \ and\ \bibinfo {author} {\bibfnamefont {G.}~\bibnamefont {Evensen}},\ }\bibfield  {title} {\enquote {\bibinfo {title} {Data assimilation in the geosciences: {{An}} overview of methods, issues, and perspectives},}\ }\href {\doibase 10.1002/wcc.535} {\bibfield  {journal} {\bibinfo  {journal} {WIREs Climate Change}\ }\textbf {\bibinfo {volume} {9}},\ \bibinfo {pages} {e535} (\bibinfo {year} {2018})}\BibitemShut {NoStop}%
\bibitem [{\citenamefont {He}\ \emph {et~al.}(2016)\citenamefont {He}, \citenamefont {Zhang}, \citenamefont {Ren},\ and\ \citenamefont {Sun}}]{he2016deep}%
  \BibitemOpen
  \bibfield  {author} {\bibinfo {author} {\bibfnamefont {K.}~\bibnamefont {He}}, \bibinfo {author} {\bibfnamefont {X.}~\bibnamefont {Zhang}}, \bibinfo {author} {\bibfnamefont {S.}~\bibnamefont {Ren}}, \ and\ \bibinfo {author} {\bibfnamefont {J.}~\bibnamefont {Sun}},\ }\bibfield  {title} {\enquote {\bibinfo {title} {Deep residual learning for image recognition},}\ }in\ \href@noop {} {\emph {\bibinfo {booktitle} {Proceedings of the {{IEEE}} Conference on Computer Vision and Pattern Recognition}}}\ (\bibinfo {year} {2016})\ pp.\ \bibinfo {pages} {770--778}\BibitemShut {NoStop}%
\bibitem [{\citenamefont {Ronneberger}, \citenamefont {Fischer},\ and\ \citenamefont {Brox}(2015)}]{ronnebergerUNetConvolutionalNetworks2015}%
  \BibitemOpen
  \bibfield  {author} {\bibinfo {author} {\bibfnamefont {O.}~\bibnamefont {Ronneberger}}, \bibinfo {author} {\bibfnamefont {P.}~\bibnamefont {Fischer}}, \ and\ \bibinfo {author} {\bibfnamefont {T.}~\bibnamefont {Brox}},\ }\href {\doibase 10.48550/arXiv.1505.04597} {\enquote {\bibinfo {title} {U-{{Net}}: {{Convolutional Networks}} for {{Biomedical Image Segmentation}}},}\ } (\bibinfo {year} {2015}),\ \Eprint {http://arxiv.org/abs/1505.04597} {arxiv:1505.04597 [cs]} \BibitemShut {NoStop}%
\bibitem [{\citenamefont {Leinonen}\ \emph {et~al.}(2023)\citenamefont {Leinonen}, \citenamefont {Hamann}, \citenamefont {Nerini}, \citenamefont {Germann},\ and\ \citenamefont {Franch}}]{leinonenLatentDiffusionModels2023}%
  \BibitemOpen
  \bibfield  {author} {\bibinfo {author} {\bibfnamefont {J.}~\bibnamefont {Leinonen}}, \bibinfo {author} {\bibfnamefont {U.}~\bibnamefont {Hamann}}, \bibinfo {author} {\bibfnamefont {D.}~\bibnamefont {Nerini}}, \bibinfo {author} {\bibfnamefont {U.}~\bibnamefont {Germann}}, \ and\ \bibinfo {author} {\bibfnamefont {G.}~\bibnamefont {Franch}},\ }\href {\doibase 10.48550/arXiv.2304.12891} {\enquote {\bibinfo {title} {Latent diffusion models for generative precipitation nowcasting with accurate uncertainty quantification},}\ } (\bibinfo {year} {2023}),\ \Eprint {http://arxiv.org/abs/2304.12891} {arxiv:2304.12891 [physics]} \BibitemShut {NoStop}%
\bibitem [{\citenamefont {Rahaman}\ \emph {et~al.}(2019)\citenamefont {Rahaman}, \citenamefont {Baratin}, \citenamefont {Arpit}, \citenamefont {Draxler}, \citenamefont {Lin}, \citenamefont {Hamprecht}, \citenamefont {Bengio},\ and\ \citenamefont {Courville}}]{rahamanSpectralBiasNeural2019}%
  \BibitemOpen
  \bibfield  {author} {\bibinfo {author} {\bibfnamefont {N.}~\bibnamefont {Rahaman}}, \bibinfo {author} {\bibfnamefont {A.}~\bibnamefont {Baratin}}, \bibinfo {author} {\bibfnamefont {D.}~\bibnamefont {Arpit}}, \bibinfo {author} {\bibfnamefont {F.}~\bibnamefont {Draxler}}, \bibinfo {author} {\bibfnamefont {M.}~\bibnamefont {Lin}}, \bibinfo {author} {\bibfnamefont {F.}~\bibnamefont {Hamprecht}}, \bibinfo {author} {\bibfnamefont {Y.}~\bibnamefont {Bengio}}, \ and\ \bibinfo {author} {\bibfnamefont {A.}~\bibnamefont {Courville}},\ }\bibfield  {title} {\enquote {\bibinfo {title} {On the {{Spectral Bias}} of {{Neural Networks}}},}\ }in\ \href@noop {} {\emph {\bibinfo {booktitle} {Proceedings of the 36th {{International Conference}} on {{Machine Learning}}}}}\ (\bibinfo  {publisher} {{PMLR}},\ \bibinfo {year} {2019})\ pp.\ \bibinfo {pages} {5301--5310}\BibitemShut {NoStop}%
\bibitem [{\citenamefont {Hamill}(2001)}]{hamillInterpretationRankHistograms2001}%
  \BibitemOpen
  \bibfield  {author} {\bibinfo {author} {\bibfnamefont {T.~M.}\ \bibnamefont {Hamill}},\ }\bibfield  {title} {\enquote {\bibinfo {title} {Interpretation of {{Rank Histograms}} for {{Verifying Ensemble Forecasts}}},}\ }\href {\doibase 10.1175/1520-0493(2001)129<0550:IORHFV>2.0.CO;2} {\bibfield  {journal} {\bibinfo  {journal} {Monthly Weather Review}\ }\textbf {\bibinfo {volume} {129}},\ \bibinfo {pages} {550--560} (\bibinfo {year} {2001})}\BibitemShut {NoStop}%
\bibitem [{\citenamefont {Monache}\ \emph {et~al.}(2013)\citenamefont {Monache}, \citenamefont {Eckel}, \citenamefont {Rife}, \citenamefont {Nagarajan},\ and\ \citenamefont {Searight}}]{monacheProbabilisticWeatherPrediction2013}%
  \BibitemOpen
  \bibfield  {author} {\bibinfo {author} {\bibfnamefont {L.~D.}\ \bibnamefont {Monache}}, \bibinfo {author} {\bibfnamefont {F.~A.}\ \bibnamefont {Eckel}}, \bibinfo {author} {\bibfnamefont {D.~L.}\ \bibnamefont {Rife}}, \bibinfo {author} {\bibfnamefont {B.}~\bibnamefont {Nagarajan}}, \ and\ \bibinfo {author} {\bibfnamefont {K.}~\bibnamefont {Searight}},\ }\bibfield  {title} {\enquote {\bibinfo {title} {Probabilistic {{Weather Prediction}} with an {{Analog Ensemble}}},}\ }\href {\doibase 10.1175/MWR-D-12-00281.1} {\bibfield  {journal} {\bibinfo  {journal} {Monthly Weather Review}\ }\textbf {\bibinfo {volume} {141}},\ \bibinfo {pages} {3498--3516} (\bibinfo {year} {2013})}\BibitemShut {NoStop}%
\bibitem [{\citenamefont {Ravuri}\ \emph {et~al.}(2021)\citenamefont {Ravuri}, \citenamefont {Lenc}, \citenamefont {Willson}, \citenamefont {Kangin}, \citenamefont {Lam}, \citenamefont {Mirowski}, \citenamefont {Fitzsimons}, \citenamefont {Athanassiadou}, \citenamefont {Kashem}, \citenamefont {Madge}, \citenamefont {Prudden}, \citenamefont {Mandhane}, \citenamefont {Clark}, \citenamefont {Brock}, \citenamefont {Simonyan}, \citenamefont {Hadsell}, \citenamefont {Robinson}, \citenamefont {Clancy}, \citenamefont {Arribas},\ and\ \citenamefont {Mohamed}}]{ravuriSkilfulPrecipitationNowcasting2021}%
  \BibitemOpen
  \bibfield  {author} {\bibinfo {author} {\bibfnamefont {S.}~\bibnamefont {Ravuri}}, \bibinfo {author} {\bibfnamefont {K.}~\bibnamefont {Lenc}}, \bibinfo {author} {\bibfnamefont {M.}~\bibnamefont {Willson}}, \bibinfo {author} {\bibfnamefont {D.}~\bibnamefont {Kangin}}, \bibinfo {author} {\bibfnamefont {R.}~\bibnamefont {Lam}}, \bibinfo {author} {\bibfnamefont {P.}~\bibnamefont {Mirowski}}, \bibinfo {author} {\bibfnamefont {M.}~\bibnamefont {Fitzsimons}}, \bibinfo {author} {\bibfnamefont {M.}~\bibnamefont {Athanassiadou}}, \bibinfo {author} {\bibfnamefont {S.}~\bibnamefont {Kashem}}, \bibinfo {author} {\bibfnamefont {S.}~\bibnamefont {Madge}}, \bibinfo {author} {\bibfnamefont {R.}~\bibnamefont {Prudden}}, \bibinfo {author} {\bibfnamefont {A.}~\bibnamefont {Mandhane}}, \bibinfo {author} {\bibfnamefont {A.}~\bibnamefont {Clark}}, \bibinfo {author} {\bibfnamefont {A.}~\bibnamefont {Brock}}, \bibinfo {author} {\bibfnamefont {K.}~\bibnamefont {Simonyan}}, \bibinfo {author} {\bibfnamefont {R.}~\bibnamefont
  {Hadsell}}, \bibinfo {author} {\bibfnamefont {N.}~\bibnamefont {Robinson}}, \bibinfo {author} {\bibfnamefont {E.}~\bibnamefont {Clancy}}, \bibinfo {author} {\bibfnamefont {A.}~\bibnamefont {Arribas}}, \ and\ \bibinfo {author} {\bibfnamefont {S.}~\bibnamefont {Mohamed}},\ }\bibfield  {title} {\enquote {\bibinfo {title} {Skilful precipitation nowcasting using deep generative models of radar},}\ }\href {\doibase 10.1038/s41586-021-03854-z} {\bibfield  {journal} {\bibinfo  {journal} {Nature}\ }\textbf {\bibinfo {volume} {597}},\ \bibinfo {pages} {672--677} (\bibinfo {year} {2021})}\BibitemShut {NoStop}%
\bibitem [{\citenamefont {Raissi}, \citenamefont {Perdikaris},\ and\ \citenamefont {Karniadakis}(2019)}]{raissiPhysicsinformedNeuralNetworks2019}%
  \BibitemOpen
  \bibfield  {author} {\bibinfo {author} {\bibfnamefont {M.}~\bibnamefont {Raissi}}, \bibinfo {author} {\bibfnamefont {P.}~\bibnamefont {Perdikaris}}, \ and\ \bibinfo {author} {\bibfnamefont {G.~E.}\ \bibnamefont {Karniadakis}},\ }\bibfield  {title} {\enquote {\bibinfo {title} {Physics-informed neural networks: {{A}} deep learning framework for solving forward and inverse problems involving nonlinear partial differential equations},}\ }\href {\doibase 10.1016/j.jcp.2018.10.045} {\bibfield  {journal} {\bibinfo  {journal} {Journal of Computational Physics}\ }\textbf {\bibinfo {volume} {378}},\ \bibinfo {pages} {686--707} (\bibinfo {year} {2019})}\BibitemShut {NoStop}%
\bibitem [{\citenamefont {Meng}\ \emph {et~al.}(2022)\citenamefont {Meng}, \citenamefont {Rombach}, \citenamefont {Gao}, \citenamefont {Kingma}, \citenamefont {Ermon}, \citenamefont {Ho},\ and\ \citenamefont {Salimans}}]{mengDistillationGuidedDiffusion2022}%
  \BibitemOpen
  \bibfield  {author} {\bibinfo {author} {\bibfnamefont {C.}~\bibnamefont {Meng}}, \bibinfo {author} {\bibfnamefont {R.}~\bibnamefont {Rombach}}, \bibinfo {author} {\bibfnamefont {R.}~\bibnamefont {Gao}}, \bibinfo {author} {\bibfnamefont {D.~P.}\ \bibnamefont {Kingma}}, \bibinfo {author} {\bibfnamefont {S.}~\bibnamefont {Ermon}}, \bibinfo {author} {\bibfnamefont {J.}~\bibnamefont {Ho}}, \ and\ \bibinfo {author} {\bibfnamefont {T.}~\bibnamefont {Salimans}},\ }\href {\doibase 10.48550/arXiv.2210.03142} {\enquote {\bibinfo {title} {On {{Distillation}} of {{Guided Diffusion Models}}},}\ } (\bibinfo {year} {2022}),\ \Eprint {http://arxiv.org/abs/2210.03142} {arxiv:2210.03142 [cs]} \BibitemShut {NoStop}%
\bibitem [{\citenamefont {Hunter}(2007)}]{hunterMatplotlib2DGraphics2007}%
  \BibitemOpen
  \bibfield  {author} {\bibinfo {author} {\bibfnamefont {J.~D.}\ \bibnamefont {Hunter}},\ }\bibfield  {title} {\enquote {\bibinfo {title} {Matplotlib: {{A 2D Graphics Environment}}},}\ }\href {\doibase 10.1109/MCSE.2007.55} {\bibfield  {journal} {\bibinfo  {journal} {Computing in Science Engineering}\ }\textbf {\bibinfo {volume} {9}},\ \bibinfo {pages} {90--95} (\bibinfo {year} {2007})}\BibitemShut {NoStop}%
\bibitem [{\citenamefont {Jeong}\ and\ \citenamefont {Hussain}(1995)}]{jeongIdentificationVortex1995}%
  \BibitemOpen
  \bibfield  {author} {\bibinfo {author} {\bibfnamefont {J.}~\bibnamefont {Jeong}}\ and\ \bibinfo {author} {\bibfnamefont {F.}~\bibnamefont {Hussain}},\ }\bibfield  {title} {\enquote {\bibinfo {title} {On the identification of a vortex},}\ }\href {\doibase 10.1017/S0022112095000462} {\bibfield  {journal} {\bibinfo  {journal} {Journal of Fluid Mechanics}\ }\textbf {\bibinfo {volume} {285}},\ \bibinfo {pages} {69--94} (\bibinfo {year} {1995})}\BibitemShut {NoStop}%
\bibitem [{\citenamefont {Papoulis}\ and\ \citenamefont {Unnikrishna~Pillai}(2002)}]{papoulis2002probability}%
  \BibitemOpen
  \bibfield  {author} {\bibinfo {author} {\bibfnamefont {A.}~\bibnamefont {Papoulis}}\ and\ \bibinfo {author} {\bibfnamefont {S.}~\bibnamefont {Unnikrishna~Pillai}},\ }\href@noop {} {\emph {\bibinfo {title} {Probability, Random Variables and Stochastic Processes}}}\ (\bibinfo {year} {2002})\BibitemShut {NoStop}%
\bibitem [{\citenamefont {Ledig}\ \emph {et~al.}(2017)\citenamefont {Ledig}, \citenamefont {Theis}, \citenamefont {Huszar}, \citenamefont {Caballero}, \citenamefont {Cunningham}, \citenamefont {Acosta}, \citenamefont {Aitken}, \citenamefont {Tejani}, \citenamefont {Totz}, \citenamefont {Wang},\ and\ \citenamefont {Shi}}]{Ledig_2017_CVPR}%
  \BibitemOpen
  \bibfield  {author} {\bibinfo {author} {\bibfnamefont {C.}~\bibnamefont {Ledig}}, \bibinfo {author} {\bibfnamefont {L.}~\bibnamefont {Theis}}, \bibinfo {author} {\bibfnamefont {F.}~\bibnamefont {Huszar}}, \bibinfo {author} {\bibfnamefont {J.}~\bibnamefont {Caballero}}, \bibinfo {author} {\bibfnamefont {A.}~\bibnamefont {Cunningham}}, \bibinfo {author} {\bibfnamefont {A.}~\bibnamefont {Acosta}}, \bibinfo {author} {\bibfnamefont {A.}~\bibnamefont {Aitken}}, \bibinfo {author} {\bibfnamefont {A.}~\bibnamefont {Tejani}}, \bibinfo {author} {\bibfnamefont {J.}~\bibnamefont {Totz}}, \bibinfo {author} {\bibfnamefont {Z.}~\bibnamefont {Wang}}, \ and\ \bibinfo {author} {\bibfnamefont {W.}~\bibnamefont {Shi}},\ }\bibfield  {title} {\enquote {\bibinfo {title} {Photo-realistic single image super-resolution using a generative adversarial network},}\ }in\ \href@noop {} {\emph {\bibinfo {booktitle} {Proceedings of the {{IEEE}} Conference on Computer Vision and Pattern Recognition ({{CVPR}})}}}\ (\bibinfo {year}
  {2017})\BibitemShut {NoStop}%
\end{thebibliography}%

\end{document}